\RequirePackage{ifpdf}
\ifpdf 
\documentclass[pdftex]{sigma}
\else
\documentclass{sigma}
\fi

\newtheorem{prop}{Proposition}[section]

\newcommand{\tr}{\mathop{\rm tr}}

\newcommand{\bra}[1]{\langle{#1}|}
\newcommand{\ket}[1]{|{#1}\rangle}

\let\qed=\cqfd

\def\pour#1{_{\,\vrule height 13pt depth 1pt\> {#1}\!}}

\def\tr{\operatorname{tr}}

\def\End{\operatorname{End}}

\def\tens{\otimes}

\def\pour#1{_{\,\vrule height 13pt depth 1pt\> {#1}\!}}

\def\sul{\sum\limits}
\def\pl{\prod\limits}

\def\build#1_#2^#3{\mathrel{
\mathop{\kern 0pt#1}\limits_{#2}^{#3}}}

\newcommand{\eH}{{\EuScript{H}}}

\newcommand{\Cset }{{\mathbb C}}

\def\cK{\mathcal{K}}
\def\cB{\mathcal{B}}
\def\cH{\mathcal{H}} 

\def\cR{\mathcal{R}}

\def\cC{\mathcal{C}}
\def\cU{\mathcal{U}}

\def\cT{\mathcal{T}}

\def\cV{\mathcal{V}}

\def\cL{\mathcal{L}}

\usepackage{euscript}

\def\sg{\sigma}
\def\la{\lambda}

\def\sul{\sum\limits}
\def\pl{\prod\limits}
\def\lt({\left(}
\def\rt){\right)}

\def\tr{\operatorname{tr}}

\def\det{\operatorname{det}}

\def\sg{\sigma}
\def\la{\lambda}

\def\tend{\longrightarrow}

\numberwithin{equation}{section}

\begin{document}

\allowdisplaybreaks

\renewcommand{\thefootnote}{$\star$}

\renewcommand{\PaperNumber}{012}

\FirstPageHeading

\ShortArticleName{Spin Chains with Non-Diagonal Boundaries}

\ArticleName{Spin Chains with Non-Diagonal Boundaries\\ and Trigonometric SOS Model with Ref\/lecting End\footnote{This paper is a
contribution to the Proceedings of the International Workshop ``Recent Advances in Quantum Integrable Systems''. The
full collection is available at
\href{http://www.emis.de/journals/SIGMA/RAQIS2010.html}{http://www.emis.de/journals/SIGMA/RAQIS2010.html}}}

\Author{Ghali FILALI~$^\dag$ and Nikolai KITANINE~$^\ddag$}

\AuthorNameForHeading{G.~Filali and N.~Kitanine}

\Address{$^\dag$~Universit\'e de Cergy-Pontoise, LPTM UMR 8089 du CNRS,\\
\hphantom{$^\dag$}~2 av. Adolphe Chauvin, 95302 Cergy-Pontoise, France}
\EmailD{\href{mailto:ghali.filali@u-cergy.fr}{ghali.filali@u-cergy.fr}}

\Address{$^\ddag$~Universit\'e de Bourgogne, Institut de Math\'ematiques de Bourgogne 		
UMR 5584 du CNRS,\\
\hphantom{$^\ddag$}~9 av. Alain Savary - B.P. 47 870,
21078 Dijon, France}
\EmailD{\href{mailto:Nikolai.Kitanine@u-bourgogne.fr}{Nikolai.Kitanine@u-bourgogne.fr}}

\ArticleDates{Received October 28, 2010, in f\/inal form January 11, 2011;  Published online January 27, 2011}

\Abstract{In this paper we consider two {\it a priori} very dif\/ferent problems: construction of the eigenstates of the spin chains with non parallel boundary magnetic f\/ields and computation of the partition function for the trigonometric  solid-on-solid (SOS) model with one ref\/lecting end and domain wall boundary conditions. We show that these two problems  are related through a  gauge transformation (so-called vertex-face transformation) and can be solved using the same dynamical ref\/lection algebras.}


\Classification{82B20; 82B23}

\renewcommand{\thefootnote}{\arabic{footnote}}
\setcounter{footnote}{0}

\section{Introduction}

In 1988 E. Sklyanin proposed a way \cite{Skl88} to generalize the algebraic Bethe ansatz  \cite{FadST79} to the open integrable systems. In particular, this approach permitted to construct the eigenstates for the quantum XXZ spin chain with external boundary magnetic f\/ields parallel to the $z$ axis (diagonal boundary terms).  General algebraic framework of this method (quantum inverse scattering method) worked also for the non-parallel boundary magnetic f\/ields, i.e.\ the double row monodromy matrix, commuting transfer matrices and the trace identities were obtained in the most general case. However, the conservation of the  $z$ component of the total spin turned out to be essential for the eigenstates construction.

It is important to mention that the algebraic Bethe ansatz  establishes a clear relation between the quantum spin chains and two-dimensional models in statistical mechanics. The periodic XXZ spin chain was solved using the transfer matrix of the six-vertex model, while the open chain with diagonal boundary terms was solved using the transfer matrix of the six-vertex model with ref\/lecting ends.

This equivalence with two-dimensional models in statistical mechanics also  turned out to be essential for the computation of the correlation functions. It was shown by V.~Korepin \cite{Kor82} that the partition function of the six-vertex model with {\it domain wall boundary conditions} (DWBC) is the key element for the study of the correlation functions. The determinant representation for this partition function found by A.~Izergin~\cite{Ize87} is crucial for the computation of the  correlation functions starting from the algebraic Bethe ansatz \cite{IzeK84,KitMT99,KitMT00}.

For the open case the corresponding partition function was computed in~\cite{Tsu98}. This representation was used f\/irst to compute the scalar products and norms of the Bethe vectors \cite{Wan02} and then to study the correlation functions of the open spin chains with external boundary magnetic f\/ields parallel to the $z$ axis   \cite{KitKMNST07,KitKMNST08}.

As we  mentioned above, the construction of the eigenstates for the spin chains with arbitrary boundary magnetic f\/ields is more complicated.
Recent advances in the study of  these systems permitted to apply the usual Bethe ansatz technique to the non-diagonal case \cite{Nep04,CaoLSW03,YanZ07}.  This case has many interesting applications, for example it can be used  to study  out-of-equilibrium systems such as asymmetric simple exclusion processes (ASEP). The Bethe ansatz solution permitted to deal with problems which were out of reach for the usual  techniques, such as the relaxation dynamics \cite{EssD05,EssD06}.

The algebraic Bethe ansatz technique developed in \cite{CaoLSW03,YanZ07} is based on a gauge transformation (so-called vertex-face transformation) which diagonalizes the boundary matrices and relates the spin chain to a trigonometric solid-on-solid (SOS) model with ref\/lecting ends.  Once again, it can be demonstrated that these two models can be described using the same algebraic structures.

The dynamical Yang--Baxter algebra proved to be the main tool to solve the SOS models~\cite{GerN84,Fel95,FelV06a,FelV06b}. In particular, it was used to compute the partition function of the SOS model with DWBC \cite{Ros09,PakRS08}. Unfortunately, the f\/inal results obtained in  these papers cannot be written as a single determinant as in the six vertex case and thus, this result cannot be easily used for the computation of the correlation functions.

In this paper we consider one possible connection between the spin chains with non-diagonal boundary terms and trigonometric SOS models in a systematic way. In the f\/irst part of our paper we review the construction of the eigenstates and we give a more simple picture than~\cite{CaoLSW03,YanZ07} using two types of SOS ref\/lection algebras. Finally, we consider the SOS model with one ref\/lecting end and we  use this algebraic framework to compute the partition functions for dif\/ferent types of  domain wall boundary conditions generalizing the results of our paper~\cite{FilK10}. The main advantage of our result is that all these partition functions can be written as a single determinant.

The main aim of this paper is to relate the trigonometric SOS model with ref\/lecting end and spin chains with non-parallel boundary magnetic f\/ields in a simple and comprehensible way and to show the algebraic structures which can be used in the future to compute the correlation functions and form factors.

It is important to note that the Bethe ansatz cannot be applied for arbitrary boundary f\/ields. It works only if the parameters satisfy certain boundary constrains. It seems that it is absolutely necessary  to have at least one relation between the boundary parameters in order to construct the Bethe ansatz. Here we consider a special case when two boundary matrices can be diagonalized simultaneously by the vertex-face  transformation. It requires two constraints~\cite{YanZ07} instead of the only one obtained initially~\cite{Nep04}. We would like to underline that this case is more complicated than the general one, as in this situation the Bethe ansatz solution is not complete (it is a degenerated situation where completeness conjecture~\cite{NepR03} does not work).

The paper is organized as follows. In  Section~\ref{section2} we review the quantum inverse scattering method for the spin chains with non-diagonal boundary terms. Following \cite{Skl88} we construct the ``vertex type'' double row monodromy matrix as well as a family of commuting transfer matrices. We show how the Hamiltonian can be obtained from these transfer matrices. In   Section~\ref{section3} we introduce the vertex-face transformation which diagonalizes  the boundary matrices and transform the vertex type monodromy matrices into ``SOS type'' objects, satisfying dynamical Yang--Baxter algebra relations. We construct two double row monodromy matrices which satisfy the dynamical ref\/lection relations or dual dynamical ref\/lection relations. In  Section~\ref{section4} we use these algebras to construct eigenstates of the spin chain with non diagonal boundary terms. Finally, in the last section we use these two ref\/lection algebras to compute the partition functions of the SOS model with ref\/lecting end and domain wall boundary conditions. Some details on the algebraic Bethe ansatz, Bethe states and relations between the two dynamical ref\/lection algebras are given in the Appendices.

\section{XXZ spin chain with non diagonal boundary}\label{section2}

We consider the open XXZ spin chain with the most general non-diagonal boundary terms
\begin{gather}
  H=  \sum_{i=1}^{N-1}\left(\sigma_{i}^{x}\sigma_{i+1}^{x}+\sigma_{i}^{y}\sigma_{i+1}^{y}+\cosh \eta \sigma_{i}^{z}\sigma_{i+1}^{z}\right)\nonumber\\
\phantom{H=}{} +\frac{\sinh \eta}{\sinh \zeta \sinh \delta}\big[-\cosh \zeta \cosh \delta \sigma_{1}^{z} +  \sinh \tau \sigma_{1}^{x} - i \cosh \tau \sigma_{1}^{y}\big]\nonumber\\
\phantom{H=}{} +\frac{\sinh \eta}{\sinh  \overline\zeta \sinh  \overline\delta}\big[-\cosh \overline{\zeta} \cosh \overline{\delta} \sigma_{N}^{z} +\sinh \overline{\tau} \sigma_{N}^{x} -i \cosh \overline{\tau} \sigma_{N}^{y}\big].\label{Hamiltonian}
\end{gather}
The Hamiltonian acts in the Hilbert space $\eH=\mathop{\otimes}_{m=1}^{N}\eH_m$, $\eH_m\sim\mathbb{C}^2$.
Two boundary magnetic
 f\/ields  are described here by six parameters $\delta$, $\overline{\delta}$, $\zeta$, $\overline{\zeta} $, $\tau$ and $\overline{\tau} $.

This Hamiltonian can be obtained using the Quantum Inverse Scattering Method (QISM)~\cite{FadST79} or more precisely its boundary version \cite{Skl88} starting from the Yang--Baxter equation and usual six-vertex trigonometric $R$-matrix.

 The key element of the QISM is the quantum $R$-matrix  $R: \Cset \tend\End(V\tens V)$, $V\sim\Cset^2$ satisfying the Yang--Baxter equation.
 \begin{gather*}
  R_{12}(\lambda_{1}-\lambda_{2}) R_{13}(\lambda_{1}-\lambda_{3}) R_{23}(\lambda_{2}-\lambda_{3})=
  R_{23}(\lambda_{2}-\lambda_{3}) R_{13}(\lambda_{1}-\lambda_{3}) R_{12}(\lambda_{1}-\lambda_{2}).
\end{gather*}

 The solution corresponding to  the XXZ spin chain is the  six-vertex $R$-matrix
\begin{gather*} 
   R(\lambda)=
  \begin{pmatrix}
        \sinh(\lambda+\eta)&0&0&0 \\
        0&\sinh \lambda&\sinh \eta&0 \\
        0&\sinh \eta&\sinh \lambda&0 \\
        0&0&0&\sinh(\lambda+\eta)
  \end{pmatrix}.
\end{gather*}

This $R$-matrix is symmetric, namely $\mathcal{P}R_{12}\mathcal{P}=R_{12}$, where  $\mathcal{P}$ is the permutation operator on $V\tens V$  and satisf\/ies the following important relations:
\begin{itemize}\itemsep=0pt
\item initial condition
\begin{gather*}
 R(0)=\sinh \eta\,  \mathcal{P},
 \end{gather*}
 \item unitarity
 \begin{gather}
\label{unitarity}
 R_{12}(\lambda)R_{21}(-\lambda)=-\sinh (\lambda-\eta)\sinh (\lambda+\eta)\, {\rm Id},
 \end{gather}
 \item $\mathbb{Z}_2$ symmetry
 \begin{gather*}
 \sigma_{1}^{y} \sigma_{2}^{y} R_{12}(\lambda) \sigma_{1}^{y} \sigma_{2}^{y}=R_{12}(\lambda),
 \end{gather*}
 \item crossing symmetry
 \begin{gather}
\label{crossing}
  - \sg^y_1  R^{t_1}_{12}(-\lambda-\eta)  \sg^y_1 = R_{21}(\lambda).
\end{gather}
\end{itemize}
Here
$R_{21}=\mathcal{P}_{12} R_{12} \mathcal{P}_{12}$, ${}^{t_1}$~denotes the matrix transposition on the f\/irst space of the tensor product, and $\sigma_{x,y,z}$ are the usual Pauli matrices.

The bulk monodromy matrix $T_{0}(\la)\in\End(V_0\tens\eH)$ of the inhomogeneous system is constructed as:
\begin{gather}\label{bulk_monodromy}
T_{0}(\lambda)=R_{01}(\lambda-\xi_{1})\cdots R_{0N}(\lambda-\xi_{N}).
\end{gather}
In these last expressions, $R_{0m}$ denotes the $R$-matrix in $\End(V_0\tens\eH_m)$, and $\xi_1,\xi_2,\ldots,\xi_N$ are arbitrary
complex parameters (inhomogeneity parameters) attached to the dif\/ferent sites of the chain of  length $N$.

It is easy to show that it satisf\/ies the Yang--Baxter algebra relations:
\begin{gather*}
R_{12}(\lambda_{1}-\lambda_{2})T_{1}(\lambda_{1})T_{2}(\lambda_{2})=T_{2}(\lambda_{2})T_{1}(\lambda_{1})R_{12}(\lambda_{1}-\lambda_{2}).
\end{gather*}

This monodromy matrix can be used to construct the eigenstates of the periodic XXZ chain~\cite{FadST79}. For the open chain a more complicated construction is needed to reconstruct the Hamiltonian and, eventually, to solve the model~\cite{Skl88}.

Taking account boundaries require the boundary $K_{\pm}$ matrix which satisfy the ref\/lection equation \cite{Che84,Skl88}:
\begin{gather}
R_{12}(\lambda_{12})K_{-}(\lambda_{1})R_{21}(\overline{\lambda}_{12})K_{-}(\lambda_{2}) =K_{-}(\lambda_{2})R_{12}(\overline{\lambda}_{12})K_{-}(\lambda_{1})R_{21}(\lambda_{12})
\label{reflection-vertex}
\end{gather}
and its dual:
\begin{gather*}
R_{12}(\lambda_{21}) K^{t_{1}}_{+}(\lambda_{1})R_{21}(-\overline{\lambda}_{12}-2\eta)K^{t_{2}}_{+}(\lambda_{2})
 =K^{t_{2}}_{+}(\lambda_{2})R_{12}(-\overline{\lambda}_{12}-2\eta)K^{t_{1}}_{+}(\lambda_{1})R_{21}({\la}_{21}),
\end{gather*}
here we used the following short notations $\la_{12}=\la_1-\la_2$, $\overline{\la}_{12}=\la_1+\la_2$.

We consider the most general solution of these equations \cite{GhoZ94}:
\begin{gather*}
   K_{-}(\lambda)  \equiv K_{-}(\lambda;\delta,\zeta,\tau) \\
   \hphantom{K_{-}(\lambda)}{}
 =
\begin{pmatrix}
\dfrac{\cosh(\delta+\zeta) e^{-\lambda}-\cosh(\delta-\zeta) e^{\lambda}}{2\sinh(\delta+\lambda)\sinh(\lambda+\zeta)}& e^{-\tau} \dfrac{\sinh(2\lambda)}{2\sinh(\delta+\lambda)\sinh(\lambda+\zeta)}\vspace{2mm}\\
-e^{\tau} \dfrac{\sinh(2\lambda)}{2\sinh(\delta+\lambda)\sinh(\lambda+\zeta)}& \dfrac{\cosh(\delta+\zeta) e^{\lambda}-\cosh(\delta-\zeta) e^{-\lambda}}{2\sinh(\delta+\lambda)\sinh(\lambda+\zeta)}
\end{pmatrix}
\end{gather*}
and  $K_{+}(\lambda)=K_{-}(-\lambda-\eta;\overline{\delta},\overline{\zeta},\overline{\tau})$. The complete inhomogeneous spin chain algebra is actually described by the double row monodromy matrix:
\begin{gather*}
({U}_{-})_{0}(\lambda)=T_{0}(\lambda)(K_{-})_{0}(\lambda)\widehat{T}_{0}(\lambda)
=\begin{pmatrix}
{A}_{-}(\lambda)&{B}_{-}(\lambda)\\
{C}_{-}(\lambda)&{D}_{-}(\lambda)
\end{pmatrix}_{[0]},
\end{gather*}
or, equivalently, by the  dual one:
\begin{gather*}
({U}^{t_{0}}_{+})_{0}(\lambda) =T_{0}^{t_{0}}(\lambda)(K^{t_{0}}_{+})_{0}(\lambda)\widehat{T}_{0}^{t_{0}}(\lambda)
=\begin{pmatrix}
{A}_{+}(\lambda)&{C}_{+}(\lambda)\\
{B}_{+}(\lambda)&{D}_{+}(\lambda)
\end{pmatrix}_{[0]},
\end{gather*}
where, due to the unitarity (\ref{unitarity}) and the crossing symmetry (\ref{crossing}):
\begin{gather*}
\widehat{T}_{0}(\lambda)   \equiv R_{N0}(\lambda+\xi_{N})\cdots R_{10}(\lambda+\xi_{1})
=\gamma(\lambda)\sigma_{0}^{y}T_{0}^{t_{0}}(-\lambda-\eta)\sigma_{0}^{y}
=\widehat{\gamma}(\lambda)T_{0}^{-1}(-\lambda),
\end{gather*}
with  normalization coef\/f\/icients:
\begin{gather*}
\gamma(\lambda)=(-1)^{N},\qquad \widehat{\gamma}(\lambda)=\gamma(\lambda)\prod_{i=1}^{N}\sinh(\lambda+\xi_{i}-\eta)\sinh(\lambda+\xi_{i}+\eta).
\end{gather*}
The double row monodromy matrix  satisf\/ies the ref\/lection algebra relation
\begin{gather*}
R_{12}(\la_{12}) ({U}_{-})_{1}(\lambda_{1})R_{21}(\overline{\la}_{12})({U}_{-})_{2}(\lambda_{2})
 =({U}_{-})_{2}(\lambda_{2})R_{12}(\overline{\la}_{12})({U}_{-})_{1}(\lambda_{1})R_{21}(\la_{12}).
\end{gather*}
The corresponding relation for  the dual double monodromy matrix  is
\begin{gather*}
R_{12}(\lambda_{21}) \big({U}_{+}^{t_{1}}\big)_{1}(\lambda_{1})R_{21}(-\overline{\la}_{12}-2\eta)\big(U_{+}^{t_{2}}\big)_{2}(\lambda_{2}) \\
\qquad
 =\big(U_{+}^{t_{2}}\big)_{2}(\lambda_{2})R_{12}(-\overline{\la}_{12}-2\eta)\big(U_{+}^{t_{1}}\big)_{1}(\lambda_{1})R_{21}(\lambda_{21}).
\end{gather*}

The transfer matrix $\mathbf{T}_{\rm XXZ}(\la)\in\End\eH$ can be constructed from both double row monodromy matrices
\begin{gather}\label{transfer}
   \mathbf{T}_{\rm XXZ}(\la) = \tr_0 \{ (K_+)_{0}(\la)\,({U}_{-})_{0}(\lambda)\}=\tr_0 \big\{ \big(K_-^{t_{0}}\big)_{0}(\la)\big(U_{+}^{t_{0}}\big)_{0}(\lambda)\big\}.
\end{gather}
Here the trace is taken over the auxiliary space $V_0$.
It was shown by Sklyanin \cite{Skl88} that the transfer matrices commute for any value of the spectral parameter,
\begin{gather*} 
\left[  \mathbf{T}_{\rm XXZ}(\la),  \mathbf{T}_{\rm XXZ}(\mu)\right]=0.
\end{gather*}
  In the homogeneous limit ($\xi_m=0$ for $m=1,\ldots,N$), the Hamiltonian of the open XXZ spin chain with most general boundary f\/ields (\ref{Hamiltonian}) can be obtained as the following derivative of the transfer matrix~(\ref{transfer}):
\begin{gather}\label{Ht}
  \cH=c_{1} \frac{d}{d \lambda}\mathbf{T}_{\rm XXZ}(\la)\pour{\la=0}+{\rm const},
\end{gather}
where
\begin{gather*}
c_{1}=-8\sinh(\delta)\sinh(\overline{\delta}-\eta)\sinh(\zeta)\sinh(\overline{\zeta}-\eta).
\end{gather*}

This procedure permits to construct a commuting family of the conserved charges. However to  construct the eigenstates in the framework of the algebraic Bethe ansatz  one needs a reference state $\ket{0}$ which is an eigenstate for the operators $A_\pm$ and $D_\pm$ and annihilated by the operators~$C_\pm$,
\begin{gather*}
C_\pm(\la)\ket{0}=0, \qquad A_\pm(\la)\ket{0}=a_{\pm}\ket{0}, \qquad D_\pm(\la)\ket{0}=d_{\pm}\ket{0}.
\end{gather*}
Evidently, for the spin chains with diagonal boundary terms a ferromagnetic state with all the spins up is a state with such properties.
 But spin chain where the total spin component~$\mathbf{S}^{z}$ is not conserved (like the periodic XYZ spin chain which is related to the elliptic 8-vertex model) do not posses a similar reference state. Indeed, to diagonalize the Hamiltonian one needs a gauge transformation which  maps the underlying vertex model into a SOS model as it was pointed out by Faddeev and Takhtadjan~\cite{FadT79}. In our case, even if the open XXZ bulk Hamiltonian still conserve this~$U(1)$ symmetry, the boundary terms break it. Thus our next goal is to  construct a gauge transformation  (vertex-face) which leads to  a SOS model with a ref\/lecting end where the boundary terms are diagonal.

\section{Vertex-face transformation}\label{section3}

The idea to use a gauge transformation to diagonalize the XXZ Hamiltonian with non-diagonal boundary terms was f\/irst applied in~\cite{CaoLSW03}, then in a more algebraic form the vertex-face transformation for this case was studied in~\cite{YanZ07}. Here we propose a simpler form for this transformation and a more explicit construction of the eigenstates.

\subsection[Dynamical Yang-Baxter algebra]{Dynamical Yang--Baxter algebra}\label{section3.1}
We introduce the following local gauge transformation
\begin{gather*}
S(\lambda;\theta)\equiv S(\lambda;\theta,\omega)=e^{\la/2}
\begin{pmatrix}
e^{-(\lambda+\theta+\omega)}&e^{-(\lambda-\theta+\omega)}\\
1&1
\end{pmatrix}.
\end{gather*}
This matrix satisf\/ies two important properties.
\begin{itemize}\itemsep=0pt
 \item It diagonalizes the boundary matrix $K_-(\la)$,
\begin{gather*}
\mathcal{K}_{-}(\lambda)=S^{-1}(\lambda;\delta-\zeta,\tau)K_{-}(\lambda)S(-\lambda;\delta-\zeta,\tau),
\end{gather*}
where the diagonal matrix $\mathcal{K}_{-}(\lambda)$ is
\begin{gather}
\mathcal{K}_{-}(\lambda)\equiv\mathcal{K}_{-}(\lambda;\delta,\zeta)
=S^{-1}(\lambda;\delta-\zeta,\tau)K_{-}(\lambda)S(-\lambda;\delta-\zeta,\tau)\nonumber\\
 \phantom{\mathcal{K}_{-}(\lambda)\equiv\mathcal{K}_{-}(\lambda;\delta,\zeta)}{}
 =\begin{pmatrix}
\dfrac{\sinh(\delta-\lambda)}{\sinh(\delta+\lambda)}&0\\
0&\dfrac{\sinh(\zeta-\lambda)}{\sinh(\zeta+\lambda)}
\end{pmatrix}.\label{k-vertexsos}
\end{gather}
\item It is a vertex-face transformation, namely:
 \begin{gather}
R_{12}(\la_{12}) S_1(\lambda_{1};\theta)S_{2}(\lambda_{2};\theta-\eta\sigma_{1}^{z})
 =S_{2}(\lambda_{2};\theta)S_{1}(\lambda_{1};\theta-\eta\sigma_{2}^{z})\mathcal{R}_{12}(\la_{12};\theta), \label{vertexsos1}
\end{gather}
where we introduced the dynamical $R$-matrix
\begin{gather}\label{DRmatrix}
   \mathcal{R}(\lambda;\theta)= \!
  \begin{pmatrix}
        \sinh (\lambda+\eta)&0&0&0 \\
        0&\dfrac{\sinh \lambda \sinh (\theta-\eta)}{\sinh \theta}&\dfrac{\sinh \eta \sinh(\theta-\lambda)}{\sinh \theta}&0 \vspace{1mm}\\
        0&\dfrac{\sinh \eta \sinh(\theta+\lambda)}{\sinh \theta}&\dfrac{\sinh \lambda \sinh (\theta+\eta)}{\sinh \theta}&0 \\
        0&0&0&\sinh (\lambda+\eta)
  \end{pmatrix}\! .\!\!\!
\end{gather}
\end{itemize}
This transformation can be understood as mapping the vertex conf\/iguration into a face con\-f\/i\-gu\-ra\-tion in the dual lattice~\cite{Bax73} and the $R$-matrix~(\ref{DRmatrix}) contains the statistical weights of the trigonometric  SOS model. It is a very well studied object generating the Felder's dynamical Yang--Baxter algebra~\cite{Fel95}. It solves the dynamical Yang--Baxter equation (DYBE),
\begin{gather*}
\mathcal{R}_{12}(\la_{12};\theta-\eta\sigma^{z}_{3}) \mathcal{R}_{13}(\lambda_{13};\theta)\mathcal{R}_{23}(\lambda_{23};\theta-\eta\sigma^{z}_{1}) = \mathcal{R}_{23}(\lambda_{23};\theta)\mathcal{R}_{13}(\lambda_{13};\theta-\eta\sigma^{z}_{2})\mathcal{R}_{12}(\lambda_{12};\theta),
\end{gather*}
and satisf\/ies the following properties which will be essential to construct the double row mo\-no\-dromy matrices:
\begin{itemize}\itemsep=0pt
\item  Ice rule
\begin{gather*}
[\sigma_{1}^{z}+\sigma_{2}^{z},\mathcal{R}_{12}(\lambda;\theta)]=0.
\end{gather*}
This symmetry is responsible of the six vertex texture of the statistical weight: $\cR_{\alpha \beta}^{\mu \nu}=0$ unless $\alpha+\beta=\mu+\nu$. It is easy to see that this relation induces a similar relation for the transposed $R$-matrix:
\begin{gather*}
[\sigma_{1}^{z}-\sigma_{2}^{z},\mathcal{R}^{t_{1}}_{12}(\lambda;\theta)]=0,
\end{gather*}
which translate into: $(\cR^{t})_{\alpha \beta}^{\mu \nu}=0$ unless $\alpha-\beta=\mu-\nu$.
\item Unitarity:
\begin{gather*}
\cR_{12}(\lambda;\theta)\cR_{21}(-\lambda;\theta)=-\sinh(\lambda-\eta)\sinh(\lambda+\eta)\,\mathrm{Id}.
\end{gather*}
\item Crossing symmetries.
The crossing relations for the dynamical $R$-matrix are not as simple as for the vertex type $R$-matrix, here we write it in the following compact form\footnote{We assume in this paper the following normal ordering: the $\sigma_{0}^{z}$ in the argument of the $\cR_{01}$ matrix (which does not commute with it) is always on the right of all other operators involved in the def\/inition of $\cR$.}:
\begin{gather}\label{dcrossing1}
-\sigma_{1}^{y}\cR_{12}^{t_{1}}(-\lambda-\eta;\theta+\eta\sigma_{1}^{z})\sigma_{1}^{y}
\frac{\sinh(\theta-\eta\sigma_{2}^{z})}{\sinh\theta}
=\cR_{21}(\lambda;\theta),
\\
\label{dcrossing2}
-\sigma_{1}^{y}\cR_{21}^{t_{1}}(-\lambda-\eta;\theta-\eta\sigma_{1}^{z})\sigma_{1}^{y}
\frac{\sinh(\theta+\eta\sigma_{2}^{z})}{\sinh\theta}
=\cR_{12}(\lambda;\theta).
\end{gather}
We should stress that these two relations are not equivalent at all since the dynamical $R$-matrix are not symmetric, but there is  a weaker condition
\item Parity:
\begin{gather}\label{parity}
\cR_{21}(\lambda;\theta)=\sigma_{1}^{x,y}\sigma_{2}^{x,y}\cR_{12}(\lambda;\theta)\sigma_{1}^{x,y}\sigma_{2}^{x,y}=\cR_{12}(\lambda;-\theta).
\end{gather}
\end{itemize}

Now we can easily introduce the dynamical monodromy matrix as an ordered product of the dynamical $R$-matrices
\begin{gather}\label{dynamicalmonodromy}
\cT_{0}(\lambda;\theta)=\cR_{01}\left(\lambda-\xi_{1};\theta-\eta\sum_{i=2}^{N}\sigma^{z}_{i}\right) \cdots \cR_{0N}(\lambda-\xi_{N};\theta).
\end{gather}
$\cT_{0}(\lambda;\theta)$ satisf\/ies the zero weight condition:
\begin{gather*}
\left[\cT_{0}(\lambda;\theta),\sigma_{0}^{z}+ \mathbf{S}^z\right]=0.
\end{gather*}
It is easy to show that this monodromy matrix satisf\/ies the dynamical Yang--Baxter relation
\begin{gather*}
\cR_{12}(\la_{12};\theta-\eta \mathbf{S}^z) \cT_{1}(\lambda_{1};\theta)\cT_{2}(\lambda_{2};\theta-\eta\sigma^{z}_{1}) = \cT_{2}(\lambda_{2};\theta)\cT_{1}(\lambda_{1};\theta-\eta\sigma_{2}^{z})\cR_{12}(\la_{12};\theta),
\end{gather*}
where the  $z$ component of the total spin $\mathbf{S}^z=\sum\limits_{i=1}^{N}\sigma^{z}_{i}$.

This dynamical monodromy matrix is related to  the usual one (\ref{bulk_monodromy}) by the vertex-face transformation $S$,
\begin{gather*}
S_{-}(\{\xi\};\theta)S_{0}(\lambda;\theta-\eta \mathbf{S}^z)\cT_{0}(\lambda;\theta)
=T_{0}(\lambda)S_{0}(\lambda;\theta)S_{-}(\{\xi\};\theta-\eta\sigma_{0}^{z}),
\end{gather*}
where
\begin{gather*}
S_{-}(\{\xi\};\theta)=S_{N}(\xi_{N};\theta)\cdots S_{1}\left(\xi_{1};\theta-\eta\sum_{i=2}^{N}\sigma^{z}_{i}\right).
\end{gather*}

We need to introduce a second  vertex-face transformation  to proceed with the construction of the double row dynamical monodromy matrix (it will be necessary to construct a dynamical analog of the~$U_+$ double row monodromy matrix in the next section).
As for the vertex case, dual algebras in dynamical context are closely related to the antipode of the corresponding dynamical quantum group, which, due to the complicated form of the crossing symmetries, is not represented by the simple matrix transposition or inversion.

The matrix $\cR(\la;\theta)$  can be also obtained by the second vertex-face transformation def\/ined by the same matrix $S$
\begin{gather*}
R_{12}(\lambda_{12})S_{2}(\lambda_{2};\theta)S_{1}(\lambda_{1};\theta
+\eta\sigma_{2}^{z})=S_{1}(\lambda_{1};\theta)S_{2}(\lambda_{2};\theta+\eta\sigma_{1}^{z})\cR_{12}(\lambda_{12};\theta),
\end{gather*}
which leads to  the following dynamical Yang--Baxter equation
\begin{gather}
\cR_{12}(\lambda_{12};\theta) \cR_{13}(\lambda_{13};\theta+\eta\sigma^{z}_{2})\cR_{23}(\lambda_{23};\theta)\nonumber \\
\qquad {}= \cR_{23}(\lambda_{23};\theta+\eta\sigma^{z}_{1})\cR_{13}(\lambda_{13};\theta)\cR_{12}(\lambda_{12};\theta+\eta\sigma^{z}_{3}).\label{DYB2}
\end{gather}

Now we can introduce the ``crossed'' $L$-operator, which will be used to construct the dynamical analog of the double row monodromy matrix $U_+$
\begin{gather*}
\cL^{t_{1}}_{12}(\lambda;\theta)=\cR_{12}^{t_{1}}(\lambda;\theta+\eta\sigma_{1}^{z})\frac{\sinh(\theta-\eta\sigma_{2}^{z})}{\sinh \theta},
\end{gather*}
and its inverse (in a convenient for us normalization)
\begin{gather*}
\widehat{\cL}^{t_{1}}_{21}(\lambda;\theta)=\cR_{21}^{t_{1}}(\lambda;\theta-\eta\sigma_{1}^{z})\frac{\sinh(\theta+\eta\sigma_{2}^{z})}{\sinh \theta}.
\end{gather*}
Due to the dynamical Yang--Baxter equation (\ref{DYB2}) and the crossing symmetry (\ref{dcrossing1}), this new $L$-operator also satisf\/ies a dynamical Yang--Baxter type equation:
\begin{gather*}
\cR_{12}({\la}_{21};\theta+ \eta\sigma_{3}^{z})\cL^{t_{1}}_{13}(\lambda_{13};\theta)\cL_{23}^{t_{2}}(\lambda_{23};\theta-\eta\sigma_{1}^{z})
 =\cL_{23}^{t_{2}}(\lambda_{23};\theta)\cL^{t_{1}}_{13}(\lambda_{13};\theta-\eta\sigma_{1}^{z})\cR_{12}({\la}_{21};\theta).
\end{gather*}

It is easy to derive the properties of this $L$-operator, such as unitarity
\begin{gather*}
\widehat{\cL}_{21}^{t_{1}}(-\lambda-\eta;\theta)\cL_{12}^{t_{1}}(\lambda-\eta;\theta)=-\sinh(\lambda-\eta)\sinh(\lambda+\eta)\,\mathrm{Id},
\end{gather*}
ice rule
\begin{gather*}
[\sigma_{1}^{z}-\sigma_{2}^{z},\cL_{12}^{t_{1}}(\lambda;\theta)]=[\sigma_{1}^{z}-\sigma_{2}^{z},\widehat{\cL}_{21}^{t_{1}}(\lambda;\theta)]=0,
\end{gather*}
and parity
\begin{gather*}
\sigma_{1}^{x,y}\sigma_{2}^{x,y}\cL_{12}^{t_{1}}(\lambda;\theta)\sigma_{1}^{x,y}\sigma_{2}^{x,y}
=\widehat{\cL}_{21}^{t_{1}}(\lambda;\theta)=\cL_{12}^{t_{1}}(\lambda;-\theta),
\end{gather*}
   from the corresponding symmetries of the dynamical $R$-matrix.

Starting from this $L$-operator we can construct a dual monodromy matrix
\begin{gather}\label{dual_mono}
\cV_{0}^{t_{0}}(\lambda;\theta)= \cL^{t_{0}}_{0 N}\left(\lambda-\xi_{N};\theta+\eta\sum_{i=1}^{N-1}\sigma_{i}^{z}\right)\cdots \cL^{t_{0}}_{0 1}(\lambda-\xi_{1};\theta).
\end{gather}
The Yang--Baxter relation for this matrix can be written in the following form
\begin{gather*}
\cR_{12}({\lambda}_{21};\theta+ \eta \mathbf{S}^z)\cV^{t_{1}}_{1}(\lambda_{1};\theta)\cV^{t_{2}}_{2}(\lambda_{2};\theta-\eta\sigma_{1}^{z})
 =\cV^{t_{2}}_{2}(\lambda_{2};\theta)\cV^{t_{1}}_{1}(\lambda_{1};\theta-\eta\sigma_{2}^{z})\cR_{12}({\la}_{21};\theta).
\end{gather*}
Once again, it is possible to construct a gauge transformation which relates this matrix to the transposed vertex monodromy matrix
\begin{gather*}
S_{+}(\{\xi\};\theta)\widetilde{S}(\lambda+\eta;\theta+\eta \mathbf{S}^z)_{0}\cV^{t_{0}}_{0}(\lambda;\theta)
=T_{0}^{t_{0}}(\lambda)\widetilde{S}(\lambda+\eta;\theta)_{0}S_{+}(\{\xi\};\theta-\eta\sigma_{0}^{z}),
\end{gather*}
where
\begin{gather*}
S_{+}(\{\xi\};\theta)=S_{1}(\xi_{1};\theta)\cdots S_{N}\left(\xi_{N};\theta+\eta\sum_{i=1}^{N-1}\sigma_{i}^{z}\right),
\end{gather*}
and
\begin{gather*}
\widetilde{S}_{0}(\lambda;\theta)=\sigma_{0}^{y}S_{0}(\lambda;\theta)\sigma_{0}^{y}.
\end{gather*}

\subsection{Dynamical ref\/lection algebra}\label{section3.2}

Now we can proceed with a construction of the double row monodromy matrix. The key ingredient is once again the ref\/lection equation. It easy to check that for  the diagonal $\mathcal{K}$ matrix (\ref{k-vertexsos}) satisfy the following dynamical ref\/lection relation for $\theta=\delta-\zeta$
\begin{gather}
\cR_{12}(\la_{12};\theta) (\mathcal{K}_{-})_{1}(\lambda_{1})\cR_{21}(\overline{\la}_{12};\theta)(\mathcal{K}_{-})_{2}(\lambda_{2})\nonumber\\
 \qquad{} = (\mathcal{K}_{-})_{2}(\lambda_{2})\cR_{12}(\overline{\la}_{12};\theta)(\mathcal{K}_{-})_{1}(\lambda_{1})\cR_{21}(\la_{12};\theta).
\label{dreflexion}
\end{gather}
This is essentially the ref\/lection equation introduced in~\cite{Che84}, with the dynamical $R$-matrix. This equation can be checked directly but also it can be derived from the vertex ref\/lection equation using the gauge transformation $S$:

\begin{prop}
The vertex type reflection equation  \eqref{reflection-vertex}  for $K_{-}$ is equivalent to the SOS type reflection equation for $\mathcal{K}_{-}$ if we the SOS parameter is chosen as $\theta=\delta-\zeta$, $\omega=\tau$.
\end{prop}

Using (\ref{vertexsos1})    the left hand side of the ref\/lection equation (\ref{reflection-vertex}) can be rewritten as
\begin{gather*}
R_{12}(\la_{12}) K_{-}(\lambda_{1})R_{21}(\overline{\la}_{12})K_{-}(\lambda_{2})  = S_{2}(\lambda_{2};\theta)S_{1}(\lambda_{1};\theta-\eta\sigma^{z}_{2})
\cR_{12}(\la_{12};\theta)\mathcal{K}_{1}(\lambda_{1})\\
\qquad{} \times \cR_{21}(\overline{\la}_{12};\theta)\mathcal{K}_{2}(\lambda_{2})
S^{-1}_{1}(-\lambda_{1};\theta-\eta\sigma^{z}_{2})S^{-1}_{2}(-\lambda_{2};\theta).
\end{gather*}
To obtain this formula we used the fact that $\mathcal{K}_{-}$ is diagonal. Similar calculation for the right hand side give us:
\begin{gather*}
K_{-}(\lambda_{2}) R_{12}(\overline{\la}_{12})K_{-}(\lambda_{1})R_{21}(\la_{12})
 = S_{2}(\lambda_{2};\theta)S_{1}(\lambda_{1};\theta-\eta\sigma^{z}_{2})\
\mathcal{K}_{2}(\lambda_{2})\cR_{12}(\overline{\la}_{12};\theta)\\
\qquad{} \times \mathcal{K}_{1}(\lambda_{1};)\cR_{21}(\la_{12};\theta)
S^{-1}_{1}(-\lambda_{1};\theta-\eta\sigma^{z}_{2})S^{-1}_{2}(-\lambda_{2};\theta).
\end{gather*}
Which f\/inally leads to the equation (\ref{dreflexion}).
In a similar way one can prove that the dynamical double row monodromy matrix def\/ined as
\begin{gather}
 (\mathcal{\cU}_{-})_{0}(\lambda;\theta)=\cT_{0}(\lambda;\theta)(\mathcal{K}_{-})_{0}(\lambda)\widehat{\cT}_{0}(\lambda;\theta) =\begin{pmatrix}
\mathcal{A}_{-}(\lambda;\theta)&\mathcal{B}_{-}(\lambda;\theta)\\
\mathcal{C}_{-}(\lambda;\theta)&\mathcal{D}_{-}(\lambda;\theta)
\end{pmatrix}_{[0]},
\label{dyndr}
\end{gather}
where
\begin{gather*}
\widehat{\cT}_{0}(\lambda;\theta)\equiv   \cR_{N0}(\lambda+\xi_{N};\theta)\cdots\cR_{10}
\left(\lambda+\xi_{1};\theta-\eta\sum_{i=2}^{N}\sigma_{i}^{z}\right) \\
\phantom{\widehat{\cT}_{0}(\lambda;\theta)}{}
= \gamma(\lambda)\sigma_{0}^{y}\cT_{0}^{t_{0}}(-\lambda-\eta;\theta+\eta\sigma_{0}^{z})\sigma_{0}^{y}
\frac{\sinh(\theta-\eta\,\mathbf{S}^z)}{\sinh(\theta)}
=  \widehat{\gamma}(\lambda)\cT_{0}^{-1}(-\lambda;\theta),
\end{gather*}
satisf\/ies the dynamical ref\/lection algebra relation,
\begin{gather}
\cR_{12}(\lambda_{12}
; \theta-\eta\, \mathbf{S}^z)
(\mathcal{U}_{-})_{1}(\lambda_{1};\theta)\cR_{21}(\overline{\lambda}_{12}
 \theta-\eta \mathbf{S}^z)
(\mathcal{U}_{-})_{2}(\lambda_{2};\theta)\nonumber\\
\qquad{} =(\mathcal{U}_{-})_{2}(\lambda_{2}; \theta)\cR_{12}(\overline{\lambda}_{12}
 ;\theta-\eta\,\mathbf{S}^z)
(\mathcal{U}_{-})_{1}(\lambda_{1};\theta)\cR_{21}(\lambda_{12}
 ;\theta-\eta\,\mathbf{S}^z).\label{SOSalgebra}
\end{gather}

This double row monodromy matrix has a clear interpretation in statistical mechanics, it is an essential tool to study the trigonometric SOS model with a ref\/lecting end. We will return to this correspondence  in the last section of this paper.

It can be easily shown that the vertex-face transformation $S$ gives a relation between this dynamical double row monodromy matrix and the usual one
\begin{gather}\label{VSOSstate1}
S_{-}(\{\xi\};\theta)S_{0}(\lambda;\theta-\eta \mathbf{S}^z)(\mathcal{U}_{-})_{0}(\lambda;\theta)=(U_{-})_{0}(\lambda)S_{-}(\{\xi\};\theta)S_{0}(-\lambda;\theta-\eta \mathbf{S}^z).
\end{gather}

\subsection{Dual ref\/lection algebra}\label{section3.3}

In the vertex case we constructed two double row monodromy matrices: one around the~$K_-$ matrix and second one around the $K_+$ matrix. The same procedure is possible in the dynamical case. The only dif\/ference is that because of a more complicated crossing relations~(\ref{dcrossing1}) and~(\ref{dcrossing2}) instead of the transposed monodromy matrix we will use here the dual monodromy matrix $\cV^{t_0}(\la)$~(\ref{dual_mono}).

We start once again with a diagonal solution $\cK_+(\la)$ of the dual ref\/lection equation with the dynamical $R$-matrix,
\begin{gather*}
\cR_{12}(\lambda_{21};\overline\theta)\big(\mathcal{K}^{t_{1}}_{+}\big)_{1}(\lambda_{1})\cR_{21}(-\overline{\lambda}_{12}-2\eta;\overline\theta)
\big(\mathcal{K}_{+}^{t_{2}}\big)_{2}(\lambda_{2})\\
\qquad{} =\big(\mathcal{K}_{+}^{t_{2}}\big)_{2}(\lambda_{2})\cR_{12}(-\overline{\lambda}_{12}-2\eta;\overline\theta)\big(\mathcal{K}_{+}^{t_{1}}\big)_{1}
(\lambda_{1})\cR_{21}(\lambda_{21};\overline\theta).
\end{gather*}
It can be expressed in terms of the $\cK_-$ matrix,
\begin{gather*}
\mathcal{K}_{+}(\lambda) \equiv \mathcal{K}_{+}(\lambda;\overline{\delta},\overline{\zeta})
=
 \mathcal{K}_{-}(-\lambda-\eta;\overline{\delta},\overline{\zeta}).
\end{gather*}
The crucial point here is that both the vertex boundary matrix $K_+$ and the dynamical bounda\-ry~$\cK_+$ are related each other by mean of the second Vertex-face transformation:
\begin{gather*}
\mathcal{K}^{t}_{+}(\lambda)=\widetilde{S}^{-1}(\lambda+\eta; \overline\delta-\overline\zeta,\overline\tau)K^{t}_{+}(\lambda)\widetilde{S}(-\lambda-\eta; \overline\delta-\overline\zeta,\overline\tau).
\end{gather*}
Note that the dynamical parameter $\overline\theta$ should not  {\it a priori} coincide with the parameter~$\theta$ from the previous section as it is chosen here $\overline\theta=\overline\delta-\overline\zeta$ to diagonalize the boundary matrix~$K_+$.

As in the previous section we can construct the following double row monodromy matrix
\begin{gather*}
(\mathcal{U}_{+})^{t_{0}}_{0}(\lambda;\overline\theta)=\cV_{0}^{t}(\lambda;\overline\theta)
\big(\mathcal{K}^{t_{0}}_{+}\big)_{0}(\lambda)\widehat{\cV}_{0}^{t_{0}}(\lambda;\overline\theta)
 =\begin{pmatrix}
\mathcal{A}_{+}(\lambda;\overline\theta)&\mathcal{C}_{+}(\lambda;\overline\theta)\\
\mathcal{B}_{+}(\lambda;\overline\theta)&\mathcal{D}_{+}(\lambda;\overline\theta)
\end{pmatrix}_{[0]},
\end{gather*}
where
\begin{gather*}
\widehat{\cV}_{0}^{t_{0}}(\lambda;\overline\theta) \equiv\widehat{\cL}^{t_{0}}_{1 0}(\lambda+\xi_{1};\overline\theta)\cdots \widehat{\cL}^{t_{0}}_{N 0}\left(\lambda+\xi_{N};\overline\theta+\eta\sum_{i=1}^{N-1}\sigma_{i}^{z}\right)
 =\widetilde{\gamma}(\lambda)\big(\cV_{0}^{t_{0}}\big)^{-1}(-\lambda-2\eta;\overline\theta).
\end{gather*}
The normalization factor in the last equation can be easily computed using the unitarity of the operator  $\cL$,
\begin{gather*}
\widetilde{\gamma}(\lambda)=\gamma(\lambda)\prod_{i=1}^{N}\sinh(\lambda+\xi_{i})\sinh(\lambda+\xi_{i}+2\eta).
\end{gather*}

This  double row monodromy matrix satisf\/ies the following dynamical ref\/lection equation
\begin{gather}
\cR_{12}(\lambda_{21};\overline\theta+\eta \mathbf{S}^z)(\mathcal{U}^{t_{1}}_{+})_{1}(\lambda_{1};\overline\theta)\cR_{21} (-\overline\lambda_{12}-2\eta;\overline\theta+\eta\,\mathbf{S}^z))(\mathcal{U}_{+}^{t_{2}})_{2}(\lambda_{2};\overline\theta)\nonumber\\
\qquad{}=(\mathcal{U}_{+}^{t_{2}})_{2}(\lambda_{2};\overline\theta)\cR_{12}(-\overline{\lambda}_{12}-2\eta;\overline\theta+\eta\,\mathbf{S}^z)(\mathcal{U}_{+}^{t_{1}})_{1}(\lambda_{1};\overline\theta)\cR_{21}(\lambda_{21};\overline\theta+\eta\,\mathbf{S}^z).
\label{DSOSalgebra}
\end{gather}
Its relation with the corresponding vertex double row monodromy matrix can be easily written in the following form
\begin{gather*}
S_{+}(\{\xi\};\overline\theta)\widetilde{S}_{0}(\lambda+\eta;\overline\theta+ \eta\,\mathbf{S}^z) \big(\mathcal{U}^{t}_{+}\big)_{0}(\lambda;\overline\theta)
=\big(U_{+}^{t}\big)_{0}(\lambda) S_{+}(\{\xi\};\overline\theta)\widetilde{S}_{0}(-\lambda-\eta;\overline\theta+ \eta \mathbf{S}^z).
\end{gather*}

In the next section we show how to use these matrices to construct  eigenstates of the open XXZ spin chain with non-diagonal boundary terms.

\section{Algebraic Bethe ansatz}\label{section4}

Our goal is to construct eigenstates of the spin chain Hamiltonian~(\ref{Hamiltonian}) which is equivalent to the construction of eigenstates of the transfer matrix $\mathbf{T}_{\rm XXZ}(\la)$~(\ref{transfer}).  We have seen that usual algebraic Bethe ansatz procedure cannot be applied to this case as there is no simple reference state. Thus our strategy is to consider dynamical monodromy matrix instead of vertex one using the vertex-face correspondence. The advantage of this approach is the fact that this procedure allows us (under certain constraints) to diagonalize the boundary matrices, and hence, it becomes possible to use  the algebraic Bethe ansatz techniques.

Using (\ref{VSOSstate1}) the transfer matrix (\ref{transfer}) can rewritten in terms of the dynamical algebra ge\-ne\-rators
\begin{gather}
\mathbf{T}_{\rm XXZ}(\lambda)S_{-}(\{\xi\},\theta)=S_{-}(\{\xi\},\theta) \nonumber\\
\qquad{}\times
\mathrm{Tr}_{0}\big\{K_{+}(\lambda)S_{0}(\lambda; \theta-\eta \mathbf{S}^z)(\mathcal{U}_{-})_{0}(\lambda;\theta)S^{-1}_{0}(-\lambda;\theta-\eta \mathbf{S}^z)\big\}.
\label{transVIVF}
\end{gather}
This trace can be rewritten  in terms of the operator entries of the double row monodromy matrix
\begin{gather*}
\mathrm{Tr}_{0}\{K_{+}(\lambda)S_{0}(\lambda;\theta-\eta\mathbf{S}^z)(\mathcal{T}_{-})_{0}(\lambda;\theta)S^{-1}_{0}(-\lambda;\theta-\eta\mathbf{S}^z)\}\\
\qquad{}=\{S^{-1}_{0}(-\lambda;\theta-\eta\mathbf{S}^z)K_{+}(\lambda)S_{0}(\lambda;\theta-\eta\mathbf{S}^z)\}^{+}_{+}\mathcal{A}_{-}(\lambda;\theta)\\
\qquad\quad{}+\{S^{-1}_{0}(-\lambda;\theta-\eta\mathbf{S}^z)K_{+}(\lambda)S_{0}(\lambda;\theta-\eta\mathbf{S}^z)\}^{-}_{-}\mathcal{D}_{-}(\lambda;\theta)\\
\qquad\quad{} +\{S^{-1}_{0}(-\lambda;\theta-\eta\mathbf{S}^z-2\eta)K_{+}(\lambda)S_{0}(\lambda;\theta-\eta\mathbf{S}^z)\}^{-}_{+}\mathcal{B}_{-}(\lambda;\theta)\\ \qquad\quad{}+\{S^{-1}_{0}(-\lambda;\theta-\eta\mathbf{S}^z+2\eta)K_{+}(\lambda)S_{0}(\lambda;\theta-\eta\mathbf{S}^z)\}^{+}_{-}\mathcal{C}_{-}(\lambda;\theta).
\end{gather*}

The next crucial step is to restrict our analysis to a subspace with a f\/ixed~$z$ component of the total spin (in the face picture, it has nothing to do with the XXZ spin). This restriction follows from the results of~\cite{CaoLSW03,Nep04}, it was shown that the number of Bethe roots for all the eigenstates is the same. It means that in the framework of the algebraic Bethe ansatz the number of creation operators and hence the total spin should be f\/ixed. Unfortunately it also means that in this approach the algebraic Bethe ansatz cannot lead to the complete description of the eigenstates.

Thus we consider the action of the trace in (\ref{transVIVF}) on the states $\ket{\psi}$ with a f\/ixed $z$ component of the total spin
\begin{gather*}
\mathbf{S}^z \ket{\psi}=s  \ket{\psi}.
\end{gather*}

To apply the algebraic Bethe ansatz we need to diagonalize both boundary matrices simultaneously, it means that the non diagonal terms should be zero,
\begin{gather*}
\big\{S^{-1}_{0}(-\lambda;\delta-\zeta-\eta s,\tau)K_{+}(\lambda)S_{0}(\lambda;\delta-\zeta-\eta(s-2),\tau)\big\}^{-}_{+}=0,\\
\nonumber
\big\{S^{-1}_{0}(-\lambda;\delta-\zeta-\eta s,\tau)K_{+}(\lambda)S_{0}(\lambda;\delta-\zeta-\eta(s+2),\tau)\big\}^{+}_{-}=0.
\end{gather*}
These lead to two constraints on the boundary parameters:
\begin{gather}\label{condition1}
\cosh(\overline{\delta}-\overline{\zeta})=\cosh(\delta-\zeta-\eta s+\overline{\tau}-\tau-\eta),\\
\label{condition2}
\cosh(\overline{\delta}-\overline{\zeta})=\cosh(\delta-\zeta-\eta s-\overline{\tau}+\tau+\eta),
\end{gather}
which can be solved by imposing:
\begin{gather}\label{solconditions}
\overline{\tau}= \tau+\eta+ i\pi n, \qquad
\overline{\delta}-\overline{\zeta}= \delta -\zeta
-\eta s+2 i\pi m.
\end{gather}

With these constraints we obtain the following  result:
\begin{gather*}
\mathbf{T}_{\rm XXZ}(\lambda)S_{-}(\{\xi\};\theta)\ket{\psi}=S_{-}(\{\xi\};\theta)
\mathrm{Tr}_{0}(\widetilde{\mathcal{K}}_{+}(\lambda;\overline\delta,\overline\zeta)\mathcal{T}_{-}(\lambda;\theta))\ket{\psi},
\end{gather*}
where
\begin{gather*}
\widetilde{\mathcal{K}}_{+}(\lambda;\overline\delta,\overline\zeta) =\frac{\sinh (\overline\delta-\overline\zeta-\eta \sigma^{z})}{\sinh (\overline\delta-\overline\zeta)}\mathcal{K}_{-}(-\lambda-\eta;\overline\delta,\overline\zeta).
\end{gather*}

Therefore the eigenstates of the vertex transfer matrix $\mathbf{T}_{\rm XXZ}$ can be obtained from the eigenstates of the SOS transfer matrix
\begin{gather}\label{tsos1}
\mathbf{T}_{\rm SOS_{1}}(\lambda;\theta)=\mathrm{Tr}_{0}\big(\widetilde{\mathcal{K}}_{+}(\lambda;\overline\delta,\overline\zeta) \mathcal{U}_{-}(\lambda;\theta)\big).
\end{gather}

Using the ref\/lection algebra relation  (\ref{SOSalgebra}) which contains the commutation relations for the operators $\mathcal{A}_{-}(\lambda;\theta)$, $\mathcal{B}_{-}(\lambda;\theta)$, $\mathcal{C}_{-}(\lambda;\theta)$, and $\mathcal{D}_{-}(\lambda;\theta)$, we can actually construct two sets of eigenstates of $\mathbf{T}_{\rm SOS_{1}}(\lambda;\theta)$.

The advantage of the SOS picture is the existence of  the reference state $\ket{0}$  with all the spins
up, which is an eigenstate of the operators  $\mathcal{A}_{-}(\lambda;\theta)$ and  $\mathcal{D}_{-}(\lambda;\theta)$ and it is annihilated by the operator $\mathcal{C}_{-}(\lambda;\theta)$,  $\mathcal{C}_{-}(\lambda;\theta)\ket{0}=0$. Then the eigenstates are constructed using the algebraic Bethe ansatz technique:

\begin{prop}\label{PropB}
Let $\theta=\delta-\zeta$, and $\delta$, $\zeta$, $\overline\delta$ and $\overline\zeta$ satisfy the boundary constraints  \eqref{solconditions} with total spin $s$ being even if $N$ is even and odd if $N$ is odd, $|s|< N$. Then the state
\begin{gather}\label{stateB}
\ket{\psi_{-}^{1}(\{\lambda\}_M)}=\mathcal{B}_{-}(\lambda_{1};\theta)\cdots \mathcal{B}_{-}(\lambda_{M};\theta)\ket{0}  ,\qquad M=\frac{N-s}2,
\end{gather}
 is an eigenstate of $\mathbf{T}_{\rm SOS_{1}}(\mu;\theta)$ with eigenvalue
\begin{gather*}
\Lambda_{1}(\mu;\{\lambda\};\delta, \zeta, \overline\delta,\overline\zeta)
=\frac{\sinh( \overline\zeta-\mu)\sinh( \overline\delta+\mu)\sinh(\delta-\mu)\sinh(2\mu+2\eta)}{\sinh( \overline\zeta-\mu-\eta)\sinh( \overline\delta-\mu-\eta)\sinh(\delta+\mu)\sinh(2\mu+\eta)}\\
\qquad{} \times \prod_{i=1}^{M} \frac{\sinh(\mu+\lambda_{i})\sinh(\mu-\lambda_{i}-\eta)}{\sinh(\mu+\lambda_{i}
+\eta)\sinh(\mu-\lambda_{i})}\prod_{i=1}^{N}\sinh(\mu+\xi_{i}+\eta)\sinh(\mu-\xi_{i}+\eta)\\
\qquad{} +
\frac{\sinh( \overline\zeta+\mu+\eta)\sinh(\delta+\mu+\eta)\sinh(\zeta-\mu-\eta)\sinh 2\mu}{\sinh( \overline\zeta-\mu-\eta)\sinh(\delta+\mu)\sinh(\zeta+\mu)\sinh(2\mu+\eta)}\\
\qquad{}\times \prod_{i=1}^{M}\frac{\sinh(\mu+\lambda_{i}+2\eta)\sinh(\mu-\lambda_{i}+\eta)}{\sinh(\mu+\lambda_{i}
+\eta)\sinh(\mu-\lambda_{i})}\prod_{i=1}^{N}\sinh(\mu+\xi_{i})\sinh(\mu-\xi_{i}),
\end{gather*}
provided the $\lambda_{i}$, $i=1,\dots,M$ satisfy the Bethe equation:
\begin{gather}\label{BetheeqB}
y_{1}(\lambda_{i},\{\lambda\};\delta, \zeta, \overline\delta,\overline\zeta)=y_{1}(-\lambda_{i}-\eta,\{\lambda\};\delta, \zeta, \overline\delta,\overline\zeta),\qquad i=1,\dots,M,
\end{gather}
with
\begin{gather*}
y_{1}(\lambda_{i},\{\lambda\};\delta, \zeta, \overline\delta,\overline\zeta)=\sinh(\zeta+\lambda_{i})\sinh(\delta-\lambda_{i})\sinh( \overline\zeta-\lambda_{i})\sinh( \overline\delta+\lambda_{i})\\
\qquad{} \times \prod_{k \neq i}^{M}\sinh(\lambda_{i}+\lambda_{k})\sinh(\lambda_{i}-\lambda_{k}-\eta)\times \prod_{j = 1}^{N}\sinh(\lambda_{i}+\xi_{j}+\eta)\sinh(\lambda_{i}-\xi_{j}+\eta).
\end{gather*}
\end{prop}

The proof of this proposition is a standard algebraic Bethe ansatz computation, some details on the commutation relations leading to this result are given in the Appendix~\ref{appendixA}.

It is possible to construct a second set of Bethe states starting from the second reference state $\ket{\overline{0}}$  with all the spins down. It is also an eigenstate of the operators  $\mathcal{A}_{-}(\lambda;\theta)$ and  $\mathcal{D}_{-}(\lambda;\theta)$, while   $\mathcal{B}_{-}(\lambda;\theta)\ket{\overline 0}=0$.
\begin{prop}\label{PropC}
\begin{gather}\label{stateC}
\ket{\psi_{-}^{2}(\{\lambda\}_M)}=\mathcal{C}_{-}(\lambda_{1};\theta)\cdots \mathcal{C}_{-}(\lambda_{M};\theta)\ket{\overline 0}, \qquad M=\frac{N+s}2,
\end{gather}
is an eigenstate of $\mathbf{T}_{\rm SOS_{1}}(\mu;\theta)$ with eigenvalue:
\begin{gather}\label{relationeigenBC}
\Lambda_{2}(\mu;\{\lambda\};\delta, \zeta, \overline\delta,\overline\zeta)=\Lambda_{1}(\mu;\{\lambda\};\zeta,\delta, \overline\zeta, \overline\delta),
\end{gather}
provided the $\lambda_{i}$, $i=1,\dots,M$ satisfy the Bethe equations:
\begin{gather}\label{BetheeqC}
y_{2}(\lambda_{i},\{\lambda\};\delta, \zeta, \overline\delta,\overline\zeta)
=y_{2}(-\lambda_{i}-\eta,\{\lambda\};\delta, \zeta, \overline\delta,\overline\zeta), \qquad i=1,\dots,M,
\end{gather}
with
\begin{gather}\label{relationequaBC}
y_{2}(\lambda_{i},\{\lambda\};\delta, \zeta, \overline\delta,\overline\zeta)=y_{1}(\lambda_{i},\{\lambda\};\zeta,\delta, \overline\zeta, \overline\delta).
\end{gather}
\end{prop}

These two set of Bethe states are actually related each other. More details on this relations are given the Appendix~\ref{appendixB}.
Now it is easy to construct the eigenstates of the XXZ transfer matrix;
\begin{prop}\label{states-}
Let $\theta=\delta-\zeta$, and $\delta$, $\zeta$, $\overline\delta$ and $\overline\zeta$ satisfy the boundary constraints  \eqref{solconditions} with total spin $s$ being even if $N$ is even and odd if $N$ is odd, $|s|< N$. Then the states
\[S_{-}(\{\xi\};\theta)\ket{\psi_{-}^{1}(\{\lambda\}_{M_B})},\]
 with $M_B=\frac{N-s}2 $ are eigenstates of $\mathbf{T}_{\rm XXZ}(\lambda)$ provided $\la_1,\dots,\la_{M_B}$ satisfy the Bethe equations~\eqref{BetheeqB}.

A second set of Bethe states can be constructed as
\begin{gather*}
S_{-}(\{\xi\};\theta)\ket{\psi_{-}^{2}(\{\lambda\}_{M_C})},
\end{gather*}
 with $M_C=\frac{N+s}2 $, if  $\la_1,\dots,\la_{M_C}$ satisfy \eqref{BetheeqC}.
 \end{prop}

 The corresponding eigenvalues of the Hamiltonian  is then given by \eqref{Ht}
 \begin{gather*}
 E=
\sum_{j=1}^{M}\varepsilon(\la_j)
+c_1N \coth \eta,\qquad \varepsilon(\la_j)=\frac{c_1\sinh\eta}{\sinh(\la+\eta)\sinh\la}.
 \end{gather*}
 Thus we obtained a simple algebraic expression for the eigenstates of the spin chain with non-diagonal boundaries in terms of the SOS operators and vertex-face transformation~$S$. This method however has two clear limitations: f\/irst, it works in its present form only if the constraints~(\ref{solconditions}) are satisf\/ied, and second, it permits to construct only states with f\/ixed number of operators $\mathcal{B}_{-}(\lambda;\theta)$ (or $\mathcal{C}_{-}(\lambda;\theta)$) and this set of states is evidently incomplete (it is clear from the SOS picture). Thus the problem of the complete description of the spectrum remains open.

The construction of the same eigenstates can be equivalently performed starting from the $\cU_+(\la;\overline\theta)$ ref\/lection algebra.
Following the same lines as before we obtain:
\begin{gather*}
\mathbf{T}_{\rm XXZ}
(\lambda)S_{+}(\{\xi\};\overline\theta)= S_{+}(\{\xi\};\overline\theta)
 \\
\qquad{} \times\mathrm{Tr}_{0}\big\{\widetilde{S}_{0}(\lambda+\eta;\overline\theta+\eta \mathbf{S}^z)
 (\mathcal{U}^{t_{0}}_{+})_{0}(\lambda;\overline\theta)\widetilde{S}^{-1}_{0}(-\lambda-\eta;\overline\theta+\eta \mathbf{S}^z)K_{-}^{t}(\lambda)\big\}.
\end{gather*}
Again, it is possible  to write the trace in terms of the operator entries of the double row mo\-nod\-romy matrix the $\cU_+(\la,\overline\theta)$. It is easy to check that if we consider the action of this trace on the states with $z$ component of total spin $s$ and if the constraints  (\ref{solconditions}) are satisf\/ied the non-diagonal terms in this expressions vanish
\begin{gather*}
\mathbf{T}_{\rm XXZ}
(\lambda)S_{+}(\{\xi\};\theta)\ket{\psi}=S_{+}(\{\xi\};
\overline\theta){\rm Tr}_{0}\big(\mathcal{U}^{t_{0}}_{+}(\lambda;\overline\theta)\widetilde{\mathcal{K}}^{t_{0}}_{-}(\lambda;\delta,\zeta)\big)\ket{\psi},
\end{gather*}
where
\begin{gather*}
\widetilde{\mathcal{K}}_{-}(\lambda;\delta,\zeta)=\frac{\sinh (\delta-\zeta-\eta \sigma^{z})}{\sinh (\delta-\zeta)}\mathcal{K}_{+}(-\lambda-\eta;\delta,\zeta).
\end{gather*}
It means that we replace once again the vertex transfer matrix with non-diagonal boundary matrices by a SOS transfer matrix with diagonal boundaries.
\begin{gather*}
\mathbf{T}_{\rm SOS_2}
(\lambda,\overline\theta)=\mathrm{Tr}_{0}\big(\mathcal{U}^{t_{0}}_{+}(\lambda;\overline\theta)\widetilde{\mathcal{K}}^{t_{0}}_{-}(\lambda;\delta,\zeta)\big).
\end{gather*}

\begin{prop}\label{propB+}
Let $\overline\theta=\overline\delta-\overline\zeta$,  and $\delta$, $\zeta$, $\overline\delta$ and $\overline\zeta$ satisfy the boundary constraints  \eqref{solconditions} with total spin $s$ being even if $N$ is even and odd if $N$ is odd, $|s|< N$.
\begin{gather*}
\ket{\psi_{+}^{1}(\{\lambda\}_M)}=\mathcal{B}_{+}(\lambda_{1};\overline\theta)\cdots \mathcal{B}_{+}(\lambda_{M};\overline\theta)\ket{0}  ,\qquad  M=\frac{N-s}2,
\end{gather*}
is an eigenstate of  $\mathbf{T}_{\rm SOS_{2}}(\mu;\overline\theta)$ with eigenvalue
$\Lambda_{2}(\mu;\overline\delta,\overline\zeta,\delta,\zeta)$ provided the $\lambda_{i}$, $i=1,\dots,M$ satisfy  the Bethe equations
\begin{gather}\label{BetheeqB+}
y_{2}(\lambda_{i},\{\lambda\}; \overline\delta,  \overline\zeta,\delta,\zeta)=y_{2}(-\lambda_{i}-\eta,\{\lambda\}; \overline\delta,  \overline\zeta, \delta,\zeta), \qquad i=1,\dots,M.
\end{gather}
\end{prop}

\begin{prop}\label{propC+}
\begin{gather*}
\ket{\psi_{+}^{2}(\{\lambda\})}=\mathcal{C}_{+}(\lambda_{1}; \overline\theta)\cdots \mathcal{C}_{+}(\lambda_{M}; \overline\theta)\ket{\overline{0}}, \qquad  M=\frac{N+s}2,
\end{gather*}
is an eigenstate of $\mathbf{T}_{\rm SOS_{2}}(\mu; \overline\theta)$ with eigenvalue $\Lambda_{1}(\mu;\overline\delta,\overline\zeta,\delta,\zeta)$
provided the $\lambda_{i}$, $i=1,\dots,M$ satisfy the Bethe equations
\begin{gather}\label{BetheeqC+}
y_{1}(\lambda_{i},\{\lambda\}; \overline\delta,  \overline\zeta,\delta,\zeta)=y_{1}(-\lambda_{i}-\eta,\{\lambda\}; \overline\delta,  \overline\zeta, \delta,\zeta), \qquad i=1,\dots,M.
\end{gather}
\end{prop}
Finally we can once again construct the corresponding eigenstates for the XXZ transfer matrix using the vertex-face transformation $S_+$.

\begin{prop}\label{states+}
Let $\overline\theta=\overline\delta-\overline\zeta$,  and $\delta$, $\zeta$, $\overline\delta$ and $\overline\zeta$ satisfy the boundary constraints  \eqref{solconditions} with total spin $s$ being even if $N$ is even and odd if $N$ is odd, $|s|< N$. Then the states
\begin{gather*}
S_{+}(\{\xi\};\theta)\ket{\psi_{+}^{1}(\{\lambda\}_{M_B})},
\end{gather*}
 with $M_B=\frac{N-s}2 $ are eigenstates of $\mathbf{T}_{\rm XXZ}(\lambda)$ provided $\la_1,\dots,\la_{M_B}$ satisfy the Bethe equations~\eqref{BetheeqB+}.

A second set of Bethe states can be constructed as
\begin{gather*}
S_{+}(\{\xi\};\theta)\ket{\psi_{+}^{2}(\{\lambda\}_{M_C})},
\end{gather*}
 with $M_C=\frac{N+s}2 $, if  $\la_1,\dots,\la_{M_C}$ satisfy~\eqref{BetheeqC+}.
 \end{prop}

It is easy to check that
\begin{gather*}
\Lambda_1(\mu;\delta,\zeta,\overline\delta,\overline\zeta) =\Lambda_{2}(\mu;\overline\delta,\overline\zeta,\delta,\zeta), \\
y_{1}(\lambda_{i},\{\lambda\}; \delta,\zeta,\overline\delta,  \overline\zeta) =y_{2}(\lambda_{i},\{\lambda\}; \overline\delta,  \overline\zeta,\delta,\zeta).
\end{gather*}
 It means that the eigenstates constructed by the f\/irst method (Proposition~\ref{states-}) and the states constructed using the dual algebra (Proposition~\ref{states+}) are exactly the same states. However it was already shown for diagonal boundary conditions~\cite{KitKMNST07} that it is important to be able to construct the same states by two dif\/ferent methods to proceed with the computation of the scalar products and correlation functions.

\section{SOS model with ref\/lecting end}\label{section5}

In the previous section we used the trigonometric dynamical ref\/lection algebra to construct the eigenstates of the spin chains with non-diagonal boundary f\/ields. However this algebra has another clear interpretation, it describes a trigonometric solid-on-solid  (SOS)  model with one ref\/lecting end (in the same way as usual ref\/lection algebra with diagonal matrix $K$ describes a~six-vertex model with ref\/lecting end).

The SOS model is a two dimensional statistical mechanics lattice model which can be def\/ined in terms of a {\it height function}. Every square of the lattice is characterized by a height $\theta$ and  its values for two adjacent squares dif\/fer by $\eta$. There are 6 possible face conf\/igurations
\[
\begin{array}{@{}ccc}\begin{array}{ccc}
\vphantom{\sum\limits_1^2}\theta -\eta &\vline&\theta -2\eta\\
\hline
\vphantom{\sum\limits_1^2}\theta &\vline&\theta-\eta
\end{array}\quad&\quad
\begin{array}{ccc}
\vphantom{\sum\limits_1^2}\theta +\eta&\vline&\theta +2\eta\\
\hline
\vphantom{\sum\limits_1^2}\theta &\vline&\theta +\eta
\end{array}
\quad&\quad
\begin{array}{ccc}
\vphantom{\sum\limits_1^2}\theta -\eta&\vline&\theta \\
\hline
\vphantom{\sum\limits_1^2}\theta &\vline&\theta +\eta
\end{array}
\vspace{3mm}\\
\begin{array}{@{}ccc}
\vphantom{\sum\limits_1^2}\theta +\eta&\vline&\theta \\
\hline
\vphantom{\sum\limits_1^2}\theta &\vline&\theta-\eta
\end{array}\quad&\quad
\begin{array}{ccc}
\vphantom{\sum\limits_1^2}\theta +\eta&\vline&\theta \\
\hline
\vphantom{\sum\limits_1^2}\theta &\vline&\theta +\eta
\end{array}
\quad&\quad
\begin{array}{ccc}
\vphantom{\sum\limits_1^2}\theta -\eta&\vline&\theta \\
\hline
\vphantom{\sum\limits_1^2}\theta &\vline&\theta -\eta
\end{array}
\\
\end{array}
\]
and the corresponding statistical weights $R^{a b}_{c d}$ are given by the dynamical $R$-matrix
(\ref{DRmatrix}).

We consider this model with a ref\/lecting end, which means that each horizontal line makes a U-turn on the left  (right) side of the lattice. It produces two following conf\/igurations characterized by the boundary matrix  $\mathcal{K}_- (\lambda)$ ($\mathcal{K}^{t}_+ (\lambda)$):

\begin{figure}[h]
\centerline{\includegraphics{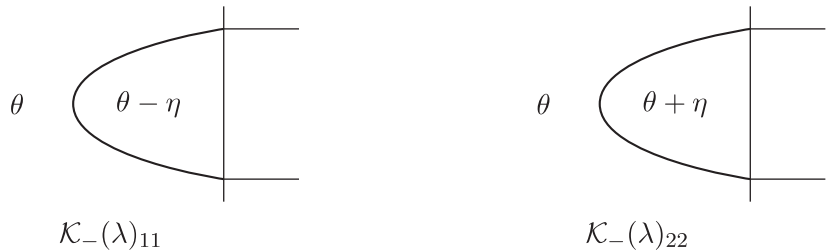}}

\caption{Boundary conf\/iguration with external height $\theta$.}
\end{figure}

It was seen in the vertex case that it is crucial to study the partition function of this model with {\it domain wall boundary conditions}. It is, in particular, necessary to study the correlation functions of the spin chains with non-diagonal boundary terms.

\begin{figure}[t]
\centering
\includegraphics[width=5.5cm]{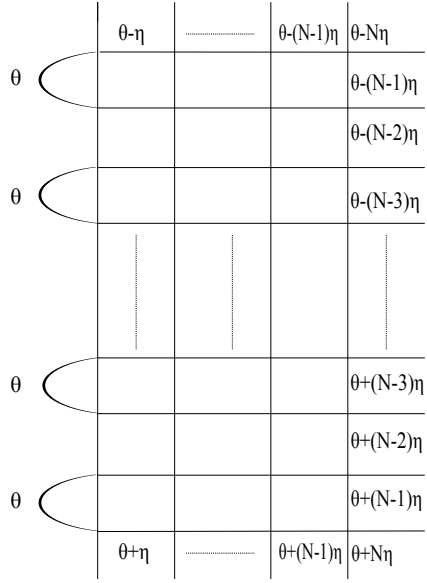}

\caption{Domain wall boundary conditions.}\label{Fig-2}
\end{figure}

These domain wall boundary conditions can be easily written for both cases: with left or right ref\/lection end, here we give the example of the left ref\/lection end (given by the matrix~$\cK_-$), see Fig.~\ref{Fig-2}.
Here the heights decrease from left to rights on the upper boundary, the heights grow from left to right on the lower boundary (we will call this situation case~I). As left external height is f\/ixed these two conditions determine completely the conf\/iguration on the right boundary (heights decreasing in the upward direction). Equivalently one can consider heights increasing on the upper boundary and decreasing on the lower one (case~II). The same two situations (case~I and case~II) are possible for the ref\/lection on the right boundary (with boundary weights given by the matrix $\cK^{t}_+$).

Thus we have four possible conf\/igurations and for all of them the partition functions can be written in terms of the double row monodromy matrix. Namely for the left ref\/lecting end the case I partition  function can be written as
\begin{gather*}
Z^{\mathcal{B_{-}}}_{N,2N}(\{\lambda\},\{\xi\},\delta,\zeta)=
\bra{\bar{0}}\prod_{i=1}^{N}
\mathcal{B}_-(\lambda_{i};\theta)  \ket{0},
\end{gather*}
while for the case II we obtain
\begin{gather*}
Z^{\mathcal{C_{-}}}_{N,2N}(\{\lambda\},\{\xi\},\delta,\zeta)=
\bra{{0}}\prod_{i=1}^{N}
\mathcal{C}_-(\lambda_{i};\theta)  \ket{\bar{0}},
\end{gather*}
where we used our standard notations $\theta=\delta-\zeta$.

 If the ref\/lecting end is on the right, we obtain the following representations:
\begin{gather*}
Z^{\mathcal{B_{+}}}_{N,2N}(\{\lambda\},\{\xi\},\overline\delta,\overline\zeta)=
\bra{\bar{0}}\prod_{i=1}^{N}
\mathcal{B}_+(\lambda_{i};\overline\theta)  \ket{0},\\ 
Z^{\mathcal{C_{+}}}_{N,2N}(\{\lambda\},\{\xi\},\overline\delta,\overline\zeta)=
\bra{{0}}\prod_{i=1}^{N}
\mathcal{C}_+(\lambda_{i};\overline\theta)  \ket{\bar{0}}.
\end{gather*}
All these partition functions can be computed following the same lines. The detailed derivation of the function  $Z^{\mathcal{B_{-}}}_{N,2N}$ is given in~\cite{FilK10}, it is shown that this function is uniquely def\/ined by the following properties
\begin{itemize}\itemsep=0pt
\item For each  parameter $\lambda_{i}$ the normalized partition function
\begin{gather*}
\tilde{Z}_{N,2N}(\{\lambda\}, \{\xi\},\delta,\zeta)=\exp\left((2N+2)\sul_{i=1}^N\lambda_{i}\right)\nonumber\\
\qquad{} \times\sinh(\delta+\lambda_i)\sinh(\zeta+\lambda_i)Z_{N,2N}^{\mathcal{B_{-}}}(\{\lambda\},\{\xi\},\delta,\zeta),
\end{gather*}
is a polynomial of degree at most  $2N+2$ in
$e^{2\lambda_{i}}$.
\item For $N=1$ the partition function is
\begin{gather*}
Z_{1,2}^{\mathcal{B_{-}}}(\lambda,\xi,\delta,\zeta)=\frac{\sinh\eta\sinh(\delta-\zeta-\eta)}{\sinh^2(\delta-\zeta)}
 \left(\frac{\sinh(\delta-\lambda)}{\sinh(\delta+\lambda)}\sinh(\lambda-\xi)\sinh(\delta-\zeta+\lambda+\xi)\right.\\
\left.
\hphantom{Z_{1,2}^{\mathcal{B_{-}}}(\lambda,\xi,\delta,\zeta)=}{}
+\frac{\sinh(\zeta-\lambda)}{\sinh(\zeta+\lambda)}\sinh(\lambda+\xi)\sinh(\delta-\zeta-\lambda+\xi)\right).
\end{gather*}
\item $Z_{N,2N}^{\mathcal{B_{-}}}(\{\lambda\},\{\xi\},\delta,\zeta)$ is symmetric in $\lambda_{i}$.
\item $Z_{N,2N}^{\mathcal{B_{-}}}(\{\lambda\},\{\xi\},\delta,\zeta)$ is symmetric in $\xi_{i}$.
\item Crossing symmetry
\begin{gather*}
Z_{N,2N}^{\mathcal{B_{-}}}(-\lambda_{i}-\eta,\{\lambda\}, \{\xi\},\delta,\zeta)
=-\frac{\sinh(2(\lambda_{i}+\eta))\sinh(\lambda_{i}+\zeta)}{\sinh(2\lambda_{i})\sinh(\lambda_{i}-\zeta+\eta)}\nonumber\\
 \qquad{} \times \frac{\sinh(\lambda_{i}+\delta)}{\sinh(\lambda_{i}-\delta+\eta)}  Z_{N,2N}^{\mathcal{B_{-}}}(\lambda_{i},\{\lambda\},\{\xi\},\delta,\zeta).
\end{gather*}
\item Recursive relations
\begin{gather*}
Z_{N,2N}^{\mathcal{B_{-}}}  (\{\lambda\},\{\xi\},\delta,\zeta)\Big|_{\la_1=\xi_1}=
\frac{\sinh\eta\sinh(\zeta-\lambda_1)}{\sinh(\zeta+\lambda_1)}
\nonumber\\
\qquad{} \times\prod_{i=1}^{N}\sinh(\lambda_{i}+\xi_{1})\frac{\sinh(\delta-\zeta+(N-2i)\eta)}{\sinh(\delta-\zeta+(N-2i+1)\eta)}\\
\qquad{} \times\prod_{i=2}^{N}\sinh(\lambda_{1}-\xi_{i}+\eta)\sinh(\lambda_{1}+\xi_{i}+\eta)\sinh(\lambda_{i}-\xi_{1}+\eta)
\\
\qquad{} \times Z_{(N-1),2(N-1)}^{\mathcal{B_{-}}}(\{\lambda\}_{2,\dots, N},\{\xi\}_{2,\dots, N},\delta,\zeta),
\\
Z_{N,2N}^{\mathcal{B_{-}}}  (\{\lambda\},\{\xi\},\delta,\zeta)\Big|_{\la_N=-\xi_1}=
\frac{\sinh\eta\sinh(\delta-\lambda_N)}{\sinh(\delta+\lambda_N)}
\\
\qquad{} \times\prod_{i=1}^{N}\sinh(\lambda_{i}-\xi_{1})\frac{\sinh(\delta-\zeta+(N-2i)\eta)}{\sinh(\delta-\zeta+(N-2i+1)\eta)}\\
\qquad{} \times\prod_{i=2}^{N}\sinh(\lambda_{N}+\xi_{i}+\eta)\sinh(\lambda_{N}-\xi_{i}+\eta)\sinh(\lambda_{i-1}+\xi_{1}+\eta) \\
\qquad{} \times Z_{(N-1),2(N-1)}^{\mathcal{B_{-}}}(\{\lambda\}_{1,\dots, N-1},\{\xi\}_{2,\dots, N},\delta,\zeta).
\end{gather*}
\end{itemize}

It means that if we f\/ind a function satisfying all these properties it is the partition function. The following proposition can be easily proved by the direct verif\/ication of these properties:

\begin{prop}
\begin{gather*}
Z_{N,2N}^{\mathcal{B_{-}}} (\{\lambda\},\{\xi\},\delta,\zeta)=\gamma(\lambda)\det M_{i j}\prod_{i=1}^{N} \left(\frac{\sinh(\delta-\zeta+\eta(N-2i)}{\sinh(\delta-\zeta+\eta(N-i))}\right) \\
\qquad{} \times\frac{\pl_{i,j=1}^{N}\sinh(\lambda_{i}+\xi_{j})\sinh(\lambda_{i}-\xi_{j})\sinh(\lambda_{i}+\xi_{j}+\eta)\sinh(\lambda_{i}-\xi_{j}+\eta)}{\pl_{1\leq i<j\leq N}\sinh(\xi_{j}+\xi_{i})\sinh(\xi_{j}-\xi_{i})\sinh(\lambda_{j}-\lambda_{i})\sinh(\lambda_{j}+\lambda_{i}+\eta)},
\end{gather*}
where
\begin{gather*}
M_{i,j}=  \frac{\sinh(\delta+\xi_j)}{\sinh(\delta+\lambda_{i})} \frac{\sinh(\zeta-\xi_j)}
{\sinh(\zeta+\lambda_{i})}   \\
\phantom{M_{i,j}=}{} \times  \frac{\sinh(2\lambda_i)\sinh\eta}{\sinh(\lambda_{i}- \xi_{j}+\eta)\sinh(\lambda_{i}+ \xi_{j}+\eta)\sinh(\lambda_{i}-\xi_{j})\sinh(\lambda_{i}+\xi_{j})}.
\end{gather*}
\end{prop}

The partition function $Z_{N,2N}^{\mathcal{C_{-}}}$ can be computed in a similar way:
 \begin{gather}\label{PartitionC-}
 Z_{N,2N}^{\mathcal{C_{-}}}(\{\lambda\},\{\xi\},\delta,\zeta)= Z_{N,2N}^{\mathcal{B_{-}}}(\{\lambda\},\{\xi\},\zeta,\delta).
 \end{gather}
It is also possible to derive it from the parity symmetry (\ref{parity}) of the dynamical $R$-matrix. More details on this are given in the Appendix~\ref{appendixB}.

 The other two partition functions can be easily obtained using the same method. It is also possible to derive them from the two previous results using an isomorphism between the dynamical ref\/lection algebras generated by the elements of~$\cU_-$ and~$\cU_+$. We give more detail on this isomorphism in  the Appendix~\ref{appendixC}. Using this isomorphism we can easily prove the following propositions:
\begin{prop}
\begin{gather*}
Z_{N}^{\mathcal{C_{+}}}(\{\lambda\},\{\xi\},\overline\delta,\overline\zeta)=Z_{N}^{\mathcal{B_{-}}}(\{\ -\lambda-\eta\},\{\xi\},\overline\delta,\overline\zeta).
\end{gather*}
\end{prop}

\begin{prop}
\begin{gather*}
Z_{N}^{\mathcal{B_{+}}}(\{\lambda\},\{\xi\},\overline\delta,\overline\zeta)=Z_{N}^{\mathcal{C_{-}}}(\{\ -\lambda-\eta\},\{\xi\},\overline\delta,\overline\zeta).
\end{gather*}
\end{prop}

Notice also that:
\begin{gather*}
Z_{N}^{\mathcal{C_{+}}}(\{\lambda\},\{\xi\},\overline\delta,\overline\zeta)=Z_{N}^{\mathcal{B_{+}}}(\{\lambda\},\{\xi\},\overline\zeta,\overline\delta).
\end{gather*}
We would like to mention that similar results can be obtained  from the so-called $F$-basis representation for the $\cB_\pm$ and $\cC_\pm$ operators. This idea was f\/irst developed in~\cite{KitMT99} for the six-vertex model. The $F$-basis for the  SOS model was constructed in~\cite{AlbBFPR00} and its application to the case with a ref\/lecting end was proposed in~\cite{YanZ10}. The computation of the partition function using this approach was recently presented by W.L.~Yang~\cite{Yan10Cairns}.

These results are interesting from  two dif\/ferent points of view. First, it is the f\/irst case where the partition function of a SOS model with domain wall boundary conditions is computed as a single determinant (and not as a sum of determinants as in~\cite{Ros09}). Second, these expressions, as it was shown in the previous sections, can be directly used to study the XXZ model with non-diagonal boundary terms. In particular, it is the f\/irst step towards the computation of the correlation functions and form-factors.

It is important to underline in conclusion that we considered here a very special case of the spin chain with non-diagonal boundaries: the boundary magnetic f\/ields are not independent but satisfy   two boundary constraints~(\ref{solconditions}). This case is interesting as it leads to a direct correspondence with a SOS model with diagonal boundary matrices. However from the spin chain point of view it is a degenerated situation which is in some sense more complicated to study than the general case with only one constraint (either condition~(\ref{condition1}) or~(\ref{condition2})). In particular, the completeness of the Bethe ansatz conjectured for the latter case~\cite{NepR03} is evidently missing in the situation considered in this paper. Thus, there are two important open problems: construction of the missing states in the case with two constraints and application of the present approach  to the most general case solvable by Bethe ansatz  with only one constraint.

\appendix

\section{Dynamical commutation relations}\label{appendixA}

In this appendix we give the derivation of the dynamical algebraic Bethe ansatz. We prove here a more general version of Proposition~\ref{PropB},   Propositions \ref{PropC}, \ref{propB+} and \ref{propC+} can be proved in a very similar way.

We consider here the following dynamical transfer matrix
\begin{gather*}
\mathbf{T}_{\rm SOS_1}(\lambda;\theta)={\rm Tr}_{0}\big(\widetilde{\mathcal{K}}_{+}(\lambda;\theta+\overline\zeta -\eta \mathbf{S}^{z},\overline\zeta) \mathcal{U}_{-}(\lambda;\theta)\big).
\end{gather*}
After our usual restriction on the subspace with the f\/ixed $z$ component of the total spin and using the boundary constraints~(\ref{solconditions}) we obtain exactly the same transfer matrix (\ref{tsos1}) as in Proposition~\ref{PropB}.

Now we can construct the eigenstates of this transfer matrix, following the usual boundary algebraic Bethe ansatz technique~\cite{Skl88}.

First it is convenient to introduce a modif\/ied operator $\widetilde{\mathcal{D}}_{-}(\lambda;\theta)$,
\begin{gather*}
\widetilde{\mathcal{D}}_{-}(\lambda;\theta) =\frac{\sinh (\theta-\eta \mathbf{S}^z+\eta)}{\sinh (\theta-\eta \mathbf{S}^z)}
\left\{\mathcal{D}_{-}(\lambda;\theta)
 -\frac{\sinh(\theta-\eta \mathbf{S}^z+2\lambda+\eta)\sinh \eta}{\sinh (2\lambda+\eta)\sinh(\theta-\eta \mathbf{S}^z+\eta)}\mathcal{A}_{-}(\lambda;\theta)\right\}.
\end{gather*}
It is easy to compute the action of  the operators $\mathcal{A}_{-}(\lambda;\theta)$, $\widetilde{\mathcal{D}}_{-}(\lambda;\theta)$ on the reference state $\ket{0}$
\begin{gather*}
\mathcal{A}_{-}(\lambda;\theta)\ket{0} =\frac{\sinh(\delta-\lambda)}{\sinh(\delta+\lambda)}\prod_{i=1}^{N}\sinh(\lambda-\xi_{i}+\eta)\sinh(\lambda+\xi_{i}+\eta)\ket{0}, \\
\widetilde{\mathcal{D}}_{-}(\lambda;\theta)\ket{0} =\frac{\sinh 2\lambda \sinh(\zeta-\lambda-\eta) \sinh(\delta+\lambda+\eta)}{\sinh(2\lambda+\eta)\sinh(\zeta+\lambda)\sinh(\delta+\lambda)} \prod_{i=1}^{N}\sinh(\lambda-\xi_{i})\sinh(\lambda+\xi_{i})|0\rangle.
\end{gather*}
The dynamical ref\/lection  relation \eqref{SOSalgebra} gives the following commutations rules for the opera\-tors~$\mathcal{A}_{-}$, $\widetilde{\mathcal{D}}_{-}$ and $\mathcal{B}_{-}$
\begin{gather*}
\mathcal{A}_{-}(\lambda_{1}; \theta)\mathcal{B}_{-}(\lambda_{2};\theta)=
 -\frac{\sinh \eta \sinh(\theta-\eta \mathbf{S}^z-2\eta-\overline\lambda_{12})}{\sinh(\theta-\eta \mathbf{S}^z-\eta)\sinh(\overline{\la}_{12}+\eta)}\mathcal{B}_{-}(\lambda_{1};\theta)\widetilde{\mathcal{D}}_{-}(\lambda_{2};\theta)\\
\phantom{\mathcal{A}_{-}(\lambda_{1}; \theta)\mathcal{B}_{-}(\lambda_{2};\theta)= }{}
 +\frac{\sinh(\overline{\la}_{12})\sinh(\la_{12}-\eta)}{\sinh(\la_{12})\sinh(\overline{\la}_{12}+\eta)}
 \mathcal{B}_{-}(\lambda_{2};\theta)\mathcal{A}_{-}(\lambda_{1};\theta)\\
\phantom{\mathcal{A}_{-}(\lambda_{1}; \theta)\mathcal{B}_{-}(\lambda_{2};\theta)= }{}
-\frac{\sinh \eta \sinh 2\lambda_{2} \sinh(\la_{12}-\theta+ \mathbf{S}^z+\eta)}{\sinh(\theta-\eta \mathbf{S}^z-\eta)\sinh(\la_{12})
\sinh(2\lambda_{2}+\eta)}\mathcal{B}_{-}(\lambda_{1};\theta)\mathcal{A}_{-}(\lambda_{2};\theta),
\\
\widetilde{\mathcal{D}}_{-}(\lambda_{1} ;\theta)\mathcal{B}_{-}(\lambda_{2};\theta)=\frac{\sinh(\overline\la_{12}+\theta-\eta \mathbf{S}^z)}{\sinh(\theta-\eta\,\mathbf{S}^z-\eta)} \\
\phantom{\widetilde{\mathcal{D}}_{-}(\lambda_{1} ;\theta)\mathcal{B}_{-}(\lambda_{2};\theta)=}{}
\times
\frac{\sinh \eta \sinh 2\lambda_{2} \sinh(2\lambda_{1}+2\eta)}{\sinh(\overline{\la}_{12}+\eta)\sinh(2\lambda_{1}+\eta)\sinh(2\lambda_{2}+\eta)}\mathcal{B}_{-}(\lambda_{1}; \theta)\mathcal{A}_{-}(\lambda_{2};\theta) \\
\phantom{\widetilde{\mathcal{D}}_{-}(\lambda_{1} ;\theta)\mathcal{B}_{-}(\lambda_{2};\theta)=}{}
 +\frac{\sinh(\la_{12}+\eta)\sinh(\overline{\la}_{12}+2\eta)}{\sinh(\la_{12})\sinh(\overline{\la}_{12}+\eta)}
 \mathcal{B}_{-}(\lambda_{2};\theta)\widetilde{\mathcal{D}}_{-}(\lambda_{1};\theta)\\
\phantom{\widetilde{\mathcal{D}}_{-}(\lambda_{1} ;\theta)\mathcal{B}_{-}(\lambda_{2};\theta)=}{}
-\frac{\sinh \eta \sinh(2(\lambda_{1}+\eta))\sinh(\la_{12}+\theta-\eta \mathbf{S}^z-\eta)}{\sinh(\la_{12})\sinh(2\lambda_{1}+\eta)\sinh(\theta-\eta \mathbf{S}^z-\eta)}\mathcal{B}_{-}(\lambda_{1};\theta)\widetilde{\mathcal{D}}_{-}(\lambda_{2};\theta).
\end{gather*}
The transfer matrix can be expressed as
\begin{gather*}
 \mathbf{T}_{\rm SOS_{1}}(\lambda;\theta)
 =\frac{\sinh(\overline\zeta+\lambda+\eta)}{\sinh(\overline\zeta-\lambda-\eta)}\widetilde{\mathcal{D}}_{-}(\lambda;\theta)  \\
 \phantom{\mathbf{T}_{\rm SOS_{1}}(\lambda;\theta)=}{}
 +\frac{\sinh(\overline\zeta-\lambda)\sinh(\overline\zeta+\theta-\eta \mathbf{S}^z+\lambda)\sinh(2\lambda+2\eta)}{\sinh(\overline\zeta-\lambda-\eta)\sinh(\overline\zeta+\theta-\eta \mathbf{S}^z-\lambda-\eta)\sinh(2\lambda+\eta)}\mathcal{A}_{-}(\lambda;\theta).
\end{gather*}
Now easily one can check using usual  algebraic Bethe ansatz  arguments that a state constructed by the action of operators $\mathcal{B}_{-}$
\begin{gather*}
\ket{\psi_{-}^{1}(\{\lambda\}_M)}=\mathcal{B}_{-}(\lambda_{1};\theta)\cdots \mathcal{B}_{-}(\lambda_{M};\theta)\ket{0},
\end{gather*}
for $0<M<N$ is an eigenstate of the transfer matrix $\mathbf{T}_{\rm SOS_1}(\mu;\theta)$ provided the spectral parameters satisfy the Bethe equations
(\ref{BetheeqB}) with $\overline\delta=\delta-\zeta+\overline\zeta-\eta(N-2M)$. Note that this result is slightly more general than  Proposition~\ref{PropB}, as we can construct eigenstates with arbitrary value of the~$z$ component of the total spin~$N-2M$. However it is important to remember that only the eigenstates with a f\/ixed value of $M$ are related to the eigenstates of the spin chains with non-diagonal boundary terms.

\section{Relations between Bethe states}\label{appendixB}

 A simple relation exist between the two set of Bethe states \eqref{stateB} and \eqref{stateC} by mean of a~ge\-ne\-ralized parity symmetry of the dynamical double row monodromy matrix~(\ref{dyndr}).
 With the help of the parity symmetries~\eqref{parity} for the dynamical $R$-matrix, we easily f\/ind the corresponding symmetry for the dynamical monodromy matrix~\eqref{dynamicalmonodromy}:
\begin{gather*}
\sigma_{0}^{x}\cT(\lambda;\theta)\sigma_{0}^{x}=\Gamma_{x}\cT(\lambda;-\theta)\Gamma_{x},
\end{gather*}
where $\Gamma_{x}=\prod\limits_{i=1}^{N}\sigma_{i}^{x}$, we use once again our notation $\theta=\delta-\zeta$. Namely, parity symmetry is equivalent to exchanging $\delta$ and $\zeta$.

A similar relation exist for the solution \eqref{k-vertexsos}:
\begin{gather*}
\sigma_{0}^{x}\mathcal{K}_{-}(\lambda;\delta,\zeta)\sigma_{0}^{x}=\mathcal{K_{-}}(\lambda;\zeta,\delta),
\end{gather*}
leading to the parity symmetry for $\mathcal{U}_{-}(\lambda;\delta-\zeta)$:
\begin{gather}\label{TRelation}
\sigma_{0}^{x}\mathcal{U}_{-}(\lambda;\delta-\zeta)\sigma_{0}^{x}=\Gamma_{x}\mathcal{U}_{-}(\lambda;\zeta-\delta)\Gamma_{x},
\end{gather}
which implies the relations between $\cB_{-}(\lambda,-\theta)$ and $\cC_{-}(\lambda,\theta)$:
\begin{gather*}
\cC_{-}(\lambda,\delta-\zeta)=\Gamma_{x}\cB_{-}(\lambda,\zeta-\delta)\Gamma_{x}.
\end{gather*}

Finally, using \eqref{TRelation}, it is obvious that:
\begin{gather}\label{GammaRelation}
\mathbf{T}_{\rm SOS_{1}}(\lambda;\theta) \equiv  \mathbf{T}_{\rm SOS_{1}}(\lambda;\delta,\zeta,\overline\delta,\overline\zeta)  =\Gamma_{x}\mathbf{T}_{\rm SOS_{1}}(\lambda;\zeta,\delta,\overline\zeta,\overline\delta)\Gamma_{x}.
\end{gather}

This last relation \eqref{GammaRelation} gives a clear understanding of the relations \eqref{relationeigenBC} and \eqref{relationequaBC} between the two set of Bethe states, their eigenvalues and Bethe equations. Namely, the two set are actually related by interchanging $\delta$ and $\zeta$. This hidden parity symmetry from the SOS point of view  ($\theta \rightarrow -\theta)$ remains obvious if we consider the Hamiltonian~\eqref{Hamiltonian} of the spin chain, which is symmetric in ($\delta$,$\zeta$) and ($\overline\delta$,$\overline\zeta$). Using this relation, we obtain a simple algebraic derivation of the relation~\eqref{PartitionC-} for the partitions functions $ Z_{N,2N}^{\mathcal{C_{-}}}$ and
$ Z_{N,2N}^{\mathcal{B_{-}}}$.

\section{Isomorphism between the dynamical algebras}\label{appendixC}

We def\/ine the following isomorphism between algebras (\ref{SOSalgebra}) and (\ref{DSOSalgebra}):
\begin{gather*}
\rho: \ \ \mathcal{T}_{-}(\lambda;\theta)\longrightarrow \Gamma_{x}\mathcal{T}_{-}^{t}(-\lambda-\eta;\theta)\Gamma_{x}.
\end{gather*}
It is easy to demonstrate that $\rho(\mathcal{T}_{-}(-\lambda-\eta;\theta))$  satisf\/ies the algebraic relations  (\ref{DSOSalgebra}).
This isomorphism tell us that,  there is a direct relations between the  two  \textit{a priori} dif\/ferent SOS models with ref\/lecting end described by~(\ref{SOSalgebra}) and~(\ref{DSOSalgebra}).   We can use this correspondence  to establish a simple relation between partitions functions. It is easy to notice that  $\mathcal{C_{+}}(\lambda;\theta)=\rho(\mathcal{B_{-}}(\lambda;\theta))=\Gamma_{x}\mathcal{B_{-}}(-\lambda-\eta;\theta)\Gamma_{x}$, hence
\begin{gather*}
Z_{N}^{\mathcal{C_{+}}}(\{\lambda\},\{\xi\},\overline{\delta},\overline{\zeta}) =\bra{0}\prod_{i=1}^{N}
\mathcal{C}_+(\lambda_{i};\overline{\theta})  \ket{\bar{0}} \nonumber 
=\bra{0}\prod_{i=1}^{N}
\Gamma_{x}\mathcal{B}_-(-\lambda_{i}-\eta;\overline{\theta}))\Gamma_{x}  \ket{\bar{0}}\\
\phantom{Z_{N}^{\mathcal{C_{+}}}(\{\lambda\},\{\xi\},\overline{\delta},\overline{\zeta})}{}
=\bra{\bar{0}}\prod_{i=1}^{N}
\mathcal{B}_-(-\lambda_{i}-\eta;\overline{\theta}) \ket{0}
=Z_{N}^{\mathcal{B_{-}}}(\{-\lambda-\eta\},\{\xi\},\overline{\delta},\overline{\zeta}).
\end{gather*}
Similar relations can be obtained for two other partitions functions. This isomorphism gives also a clear understanding of the relations between the Bethe states for the XXZ spin chain with non diagonal boundaries constructed using two algebras.

\subsection*{Acknowledgments}

The authors are grateful to the LPTHE laboratory (Universit\'e Pierre et Marie Curie, Paris 6) for hospitality which made this collaboration much easier. We would like to thank J.M.~Maillet, V.~Terras, G.~Niccoli, K.K.~Kozlowski, J.~Avan and V.~Roubtsov for many interesting discussions. N.K. is supported by the  Regional Council of Bourgundy (Conseil R\'egional de Bourgogne) FABER programm.

\pdfbookmark[1]{References}{ref}
\LastPageEnding

\end{document}